\documentclass[prx,twocolumn,superscriptaddress,longbibliography,aps]{revtex4-1}

\usepackage{bm}
\usepackage[T1]{fontenc}
\usepackage[utf8]{inputenc}
\usepackage{lmodern}
\usepackage[sumlimits]{amsmath} 
\usepackage{amssymb}
\usepackage{yhmath}
\usepackage{graphicx}
\usepackage{epstopdf}
\usepackage{latexsym}
\usepackage{color}
\usepackage{nicefrac}
\usepackage{dsfont} 
\usepackage{wasysym} 
\usepackage{stmaryrd}
\usepackage{hyperref} 
\usepackage{xcolor}
\usepackage{multirow}
\usepackage{soul}
\usepackage{adjustbox}
\usepackage{cancel}

\usepackage{esint}
\usepackage{tikz}

\definecolor{airforceblue}{rgb}{0.36, 0.54, 0.66}

\newcommand{\be}{\begin{equation}}
\newcommand{\ee}{\end{equation}}
\newcommand{\bea}{\begin{eqnarray}}
\newcommand{\eea}{\end{eqnarray}}

\usepackage{verbatim}
\usepackage{stmaryrd}
\usepackage[sumlimits]{amsmath} 
\usepackage{dsfont} 
\usepackage{caption}
\usepackage{cancel}
\usepackage{url}

\usepackage{sansmath}
\DeclareFontEncoding{LGR}{}{}
\DeclareSymbolFont{sfgreek}{LGR}{cmss}{m}{n}
\SetSymbolFont{sfgreek}{bold}{LGR}{cmss}{bx}{n}
\DeclareMathSymbol{\sxi}{\mathord}{sfgreek}{`x}
\DeclareMathSymbol{\stheta}{\mathord}{sfgreek}{`j}
\DeclareMathSymbol{\sepsilon}{\mathord}{sfgreek}{`e}
\DeclareMathSymbol{\sOmega}{\mathalpha}{sfgreek}{`W}

\newcommand{\mb}{\mathbf}
\newcommand{\bs}{\boldsymbol}
\newcommand{\mf}{\mathfrak}
\newcommand{\mc}{\mathcal}
\newcommand{\ms}{\mathsf}

\renewcommand{\ul}{\underline}

\usepackage{scalerel}

\usepackage{accsupp} 

\newcommand{\cev}[1]{\reflectbox{\ensuremath{\vec{\reflectbox{\ensuremath{#1}}}}}}
\newcommand{\shbar}{\hslash}

\usepackage{subcaption}
\usepackage{ragged2e}
\DeclareCaptionJustification{justified}{\justifying}
\begin{document}

\title{Quantum kinetic equation and thermal conductivity tensor for bosons}

\author{L\'eo Mangeolle}
\affiliation{\'{E}cole Normale Sup\'{e}rieure de
  Lyon, CNRS, Laboratoire de physique, 46, all\'{e}e
d'Italie, 69007 Lyon}
\affiliation{Kavli Institute for Theoretical Physics, University of
California, Santa Barbara, CA 93106-4030}
\author{Lucile Savary}
\affiliation{\'{E}cole Normale Sup\'{e}rieure de
  Lyon, CNRS, Laboratoire de physique, 46, all\'{e}e
d'Italie, 69007 Lyon}
\affiliation{Kavli Institute for Theoretical Physics, University of
California, Santa Barbara, CA 93106-4030}
\author{Leon Balents}
\affiliation{Kavli Institute for Theoretical Physics, University of
  California, Santa Barbara, CA 93106-4030}
\affiliation{Canadian Institute for Advanced Research, Toronto, Ontario, Canada}

\date{\today}
\begin{abstract}
We systematically derive the quantum kinetic equation in full phase
space for any quadratic hamiltonian of bosonic fields, including in
the absence of translational invariance.  This enables the treatment of boundaries, inhomogeneous systems and states with non-trivial
textures, such as skyrmions in the context of magnetic bosons. We
relate the evolution of the distribution of
bosons in phase space to single-electron, band-diagonal, physical quantities such as
Berry curvature and energy magnetization, by providing a procedure to
``diagonalize'' the Hamiltonian in phase space, using the formalism of the Moyal
product. We obtain {\em exact} equations, which can be expanded order by
order, for example in the ``smallness'' of the spatial gradients,
providing a ``semiclassical'' approximation. In turn, at first order,
we recover the usual full Boltzmann equation and give a self-contained and
exact derivation of the intrinsic thermal Hall effect of bosons.
The formulation clarifies the contribution from ``energy
magnetization'' in natural manner, and does not require the inclusion
of Luttinger's pseudo-gravitational field to obtain thermal transport quantities.
\end{abstract}

\maketitle

\section{Introduction}
\label{sec:introduction}

It is now widely recognized that the dynamics of electrons in energy
bands is fundamentally affected by topological and geometric aspects
of the Bloch eigenstates.  This is described semi-classically by Berry
phases and curvatures, which enter the equations of motion for a wave
packet \cite{sundaram1999}.  The same ideas apply to any smooth
collection of non-degenerate 
single particle states indexed by
{\em continuous} parameters, and in particular to bosonic elementary
excitations like phonons or magnons \cite{qin2012,matsumoto2014}.  These concepts are particularly
powerful in describing out of, but near equilibrium phenomena such as
charge and energy transport: incorporating the semi-classical topological
dynamics into a Boltzmann description provides the most compact
and intuitive understanding of the anomalous Hall effect, as well as
many other transport properties \cite{karplus1954hall, adams1959energy, haldane2004berry, nagaosa2006anomalous,
 sinitsyn2007semiclassical, mishchenko2014equilibrium}. 
Being semi-classical, i.e.\ allowing
simultaneous consideration of position and momentum, such a method
also makes it possible to treat inhomogeneous systems, including
boundaries, textures, etc., in a compact and natural manner. It
sometimes offers advantages even for calculation of bulk transport
properties like thermal conductivity, because it allows to transparently isolate
spurious contributions due to bound ``magnetization'' currents,
which 
have in the past obscured correct results \cite{katsura2010}.

The semi-classical approach is known to reproduce the exact result for
the intrinsic anomalous Hall effect \cite{haldane2004berry}, as is often the case for
transport coefficients. 
This is because the semi-classical
approximation is controlled by the smallness of spatial and temporal
gradients -- of the Hamiltonian and of the distribution function --
which are indeed small in a perturbative response to uniform DC fields.
Theoretically, the smallness of gradients is required {\em both} to
justify the semi-classical equations of motion, which describe the
motion of {\em one} particle in phase space, the
Boltzmann equation itself, which describes the evolution of the
{\em distribution} of particles, and which is an
approximation to the fully quantum evolution of the density matrix.
The derivation of the Boltzmann equation semi-classically from the
full quantum kinetic equation (QKE) is an old
problem studied extensively {\em prior} to the widespread
incorporation of Berry phase effects into band dynamics \cite{keldysh1964,altshuler1978,rammersmith}.  Despite the
common practice of combining these two semi-classical approximations,
they are almost always treated independently, with attention focusing
primarily on the single particle equations of motion and the Boltzmann
equation adopted without justification.

A unified derivation for the electronic case was provided by Wickles
and Belzig \cite{gosselin2007,wickles2013}.  They showed that the equation of motion
for the ``one-particle density matrix'' 
of a multi-band electron system can be systematically
reduced to the Boltzmann one, with the renormalized 
single particle
equations of motion emerging in the same treatment via a single
semi-classical expansion.  Here we extend this treatment to bosons,
which, contrary to electrons, do not have a conserved number/charge,
but only energy.  The difference of statistics, but more importantly
the latter lack of conserved charge, lead to some significant
differences from the Wickles and Belzig treatment.
Nevertheless, we obtain a full derivation of the leading semi-classical kinetic equation and associated observable quantities
such as energy and current densities, and the formulation allows a clear route to extend to higher orders in the semi-classical expansion.

The derivation we present assumes non-interacting bosons, and hence
neglects any scattering (but this can of course be added).  Aside from
the neglect of interactions, the derived quantum kinetic
equation is asymptotically exact in the limit in which the parameters of
the theory are slowly varying in space.  In particular, we obtain
exactly the thermal
conductivity tensor, which describes the heat current induced by a
small temperature gradient.  Our approach therefore
provides a self-contained and exact derivation of the intrinsic
thermal Hall effect of non-interacting bosons such as phonons and
magnons, and in the last section we show that the quantum kinetic equation (QKE) indeed is in agreement with Kubo formula
calculations \cite{qin2012,matsumoto2014}. The advantage is that the
QKE calculation is considerably more intuitive, and naturally avoids
subtleties associated with magnetization currents \cite{smrcka1977transport, cooper1997, bashan2022}, which plagued 
early Kubo calculations \cite{katsura2010}, as mentioned above. For concreteness we further
compute and plot the local (i.e.\ position-resolved) energy currents for a system of chiral phonons
(excitations of linear elasticity theory with a phonon Hall viscosity
term) in a finite geometry in the presence of a temperature gradient.

\begin{figure}[htbp]
  \includegraphics[width=\columnwidth]{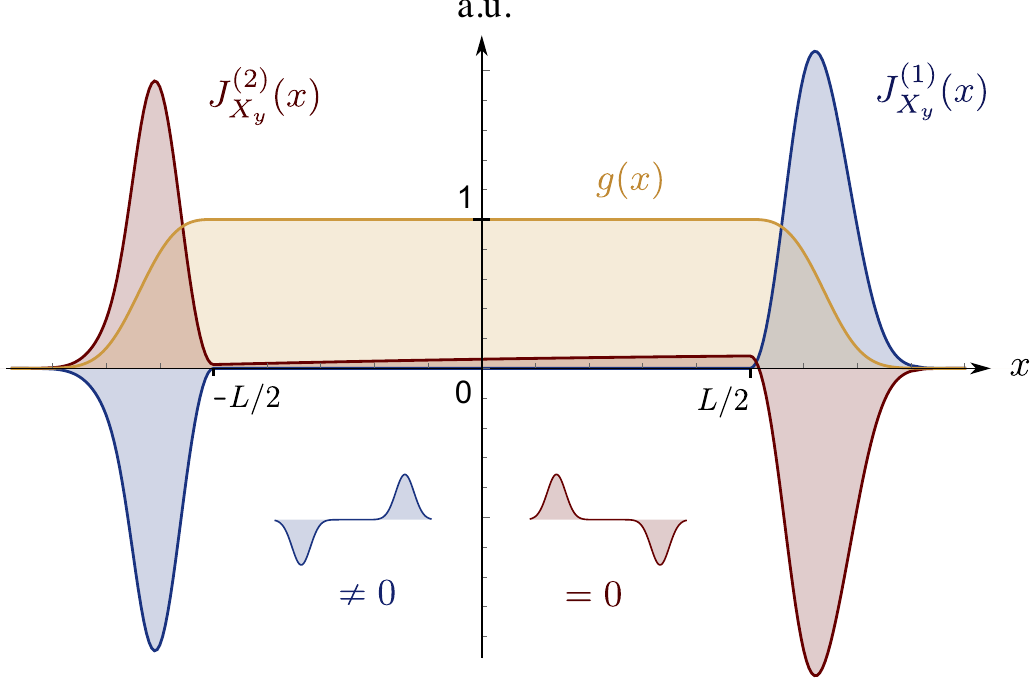}
\caption{Local momentum-integrated energy currents ${J}_{X_y}^{(1)}(x) \equiv \int_p \mathcal{J}_{X_y}^{(1)}$ (first term in
  Eq.~\eqref{eq:198}, blue) and
  ${J}_{X_y}^{(2)}(x) \equiv \int_p \mathcal{J}_{X_y}^{(2)}$ (magnetization current, second term in
  Eq.~\eqref{eq:198}, red) in arbitrary units for a system
  of chiral phonons described by the Lagrangian Eq.~\eqref{eq:98},
  calculated for $g(x)= {\rm exp} \left \{ - \frac 1 4 \Theta[|x| -
    L/2] \left (|x| - L/2 \right )^3 / \xi^3 \right \} $ (yellow,
  mimicking the existence of a boundary)
  in the presence of a constant temperature gradient $\delta T/L$,
  i.e.\ for a temperature profile $T(x)=T_0 + (x/L) \delta T$, using
  the numerical values given in Table~\ref{tab:parameter-values}.}
   \label{fig:local-currents}
\end{figure}

\section{Overview}
\label{sec:overview}

The principal result of this paper, the derivation of the
semi-classical QKE and that of observables, is somewhat technical and
subtle.  Hence we provide in this section an overview of the
derivation, highlighting the key steps, leaving full details to
Sec.~\ref{sec:detailed-derivation}.

\subsection{Basic formulation}
\label{sec:basic-formulation}

Our starting point is a general free boson system described by $2N$ free hermitian fields $\Phi_a(r)$, $r\in\mathbb{R}^d$, where $d$ is the
dimension of our system, $a=1,..,2N$.  For free fields the commutator 
\begin{equation}
  \label{eq:1}
  [{\Phi}_a(r_1),{\Phi}_b(r_2)]\equiv\hbar\,\hat{\sf \Gamma}_{ab}(r_1,r_2), 
\end{equation}
is a {\em c-number} $2N\times2N$ matrix. For example, in the context of three-dimensional elasticity, $\Phi_{1,2,3}$ might 
be the displacement fields $u_{x,y,z}$, and $\Phi_{4,5,6}$ represent
the conjugate momenta $\Pi_{x,y,z}$. 
Since the commutator is antisymmetric and the fields are
hermitian, we have $\hat{\sf \Gamma}_{ab}(r_1,r_2)=-\hat{\sf
  \Gamma}_{ba}(r_2,r_1)$ and $\hat{\sf
  \Gamma}_{ba}(r_2,r_1)=\left(\hat{\sf
    \Gamma}_{ab}(r_1,r_2)\right)^*$.

We take an arbitrary quadratic Hamiltonian
\begin{equation}
  \label{eq:2}
  \textrm{H}=\frac{1}{2}\int_{r_1,r_2}\sum_{a,b}\hat{\sf H}_{ab}(r_1,r_2)\Phi_a(r_1)\Phi_b(r_2),
\end{equation}
where $\hat{\sf H}(r_1,r_2)$ is a {\em c-number} $2N\times2N$
matrix. The hermiticity of ${\rm H}$ requires $\hat{\sf
  H}(r_2,r_1)=\left(\hat{\sf H}(r_1,r_2)\right)^*$, and to remove any
redundancy we impose $\hat{\sf H}_{ab}(r_1,r_2)=\hat{\sf
  H}_{ba}(r_2,r_1)$. Both $\hat{\sf H}$ and $\hat{\sf \Gamma}$,
regarded as matrices in the index ($a,b$) $\times$ coordinate space ($r_1,r_2$), are
hermitian matrices.

Apart from hermiticity and the symmetry/antisymmetry of
$\hat{\mathsf{H}}$/$\hat{\mathsf{\Gamma}}$, respectively, we need only
assume that $\hat{\mathsf{H}}$ has a spectrum which is bounded below,
so that the system is stable.  

The expectation value of any quadratic observable can be obtained as a linear combination of the expectation values 
\begin{equation}
  \label{eq:3}
  \hat{\sf
    F}_{ab}(r_1,r_2)\equiv\frac{1}{2}\left\langle\left\{\Phi_a(r_1),\Phi_b(r_2)\right\}\right\rangle.
\end{equation}
Here the expectation value is the expectation value with respect to
an initial many-body density matrix $\rho_0$,
$\langle A(t)\rangle={\rm Tr}[A(t)\rho_0]$, with $A(t)$ an operator in
the Heisenberg picture.  The corresponding c-number matrix
$\hat{\sf F}$, which captures any two-point expectation value of any two observables, is sometimes called `density matrix,' and is symmetric and
hermitian in the index $\times$ coordinate space.

\subsection{Transition to phase space}
\label{sec:trans-phase-space-1}

From Eq.~\eqref{eq:1} and Eq.~\eqref{eq:2} the unitary time evolution of any operator is determined.  To frame this in phase space, we introduce the Wigner transform
\begin{align}
  \label{eq:4}
  f(X,p)&=[\hat{f}]^W(X,p)\nonumber\\
 &\equiv\int_xe^{-i \frac{px}{\hbar}}\hat{f}\left(X+\frac{x}{2},X-\frac{x}{2}\right),
\end{align}
where $p,X,x\in\mathbb{R}^d$ (we can identify
$X=\frac{1}{2}(r_1+r_2)$ and $x=r_1-r_2$) and $\hbar$ is Planck's
constant divided by $2\pi$.   We denote $\int_x = \int d^dx$ for
spatial integrals, and later we will use $\int_p = \int d^dp/(2\pi\hbar)^d$ for momentum integrals (then $\int_{x,p}$ is dimensionless).    We will denote Wigner transformed matrices
by removing the hats, e.g.\ $[\hat{\sf F}]^W={\sf F}$.  These objects,
$\mathsf{F},\mathsf{H},\mathsf{\Gamma}$ etc.\ are functions on phase
space ($X,p$), and are all hermitian as matrices at fixed $X,p$.  Note
that we will use the convention that $p$ denotes momentum while $k$
denotes wavenumber, i.e.\ $p = \hbar k$.  This is useful when we keep
track of factors of $\hbar$ and to obtain the classical limit.

In terms of Wigner transformed objects, one can then derive the
equation of motion. Namely, the equation of motion for the `density matrix' in phase space is
\begin{equation}
  \label{eq:5}
     \partial_t{\sf F}(X,p)=\frac{-i}{\hbar}\left({\sf
                       K}\star{\sf F}-{\sf F}\star{\sf K}^\dagger\right),
\end{equation}
where
\begin{equation}
  \label{eq:6}
  {\sf K}(X,p)=\hbar\,{\sf \Gamma}\star{\sf H}
\end{equation}
is sometimes called the ``dynamical matrix.''  Note that $\mathsf{K}$ is {\em not} hermitian.

Here the $\star$ represents the ``star'' or ``Moyal'' product,
defined as, for any two matrix functions ${\sf O}_{1,2}$,
\begin{equation}
  \label{eq:7}
 {\sf
    O}_1\star{\sf O}_2\equiv {\sf O}_1\exp\left(i\frac{\shbar}{2}\epsilon_{\alpha\beta}\cev{\partial}_\alpha\vec{\partial}_\beta\right){\sf O}_2,
\end{equation}
where repeated indices are summed, $\alpha,\beta=1,..,2d$,
$\partial_\alpha\equiv\partial_{q_\alpha}$, where
$q_{\mu}=X_{\mu}$ and
$q_{d+\mu}=p_{\mu}$ for $\mu=1,..,d$, and where we have defined $\epsilon_{\alpha\beta}$ to be such that
\begin{equation}
  \label{eq:8}
  \epsilon_{X_\mu p_\nu}=-\epsilon_{p_\mu X_\nu}=\delta_{\mu\nu},
\end{equation}
with $\mu=1,..,d$.
Note that $\shbar=\hbar$ but we are formally distinguishing the two here
  to treat $\shbar$ as an expansion parameter when it stems from an
  expansion of Eq.~\eqref{eq:7} \cite{wickles2013}. We provide more details in
Appendix~\ref{sec:moyal-product}. See also Refs.~\cite{rammersmith,polkovnikov2010,wickles2013}.

\subsection{Semiclassical solution}
\label{sec:semicl-solut}

A proper semi-classical limit is obtained by first formally
diagonalizing the dynamical matrix and distribution function, i.e.\
reducing them to scalars within each band, and then performing a small
$\shbar$ expansion, which expands the star products Eq.~\eqref{eq:7}.  Diagonalization is accomplished via a similarity
transformation $\mathsf{S}$ such that $\mathsf{S}^{-1} \star
\mathsf{K} \star \mathsf{S} = \mathsf{K}_d$ is a diagonal matrix of
mode frequencies.  Similarly, $\mathsf{F} = \mathsf{S} \star
\mathsf{F}_d \star \mathsf{S}^\dagger$, with $\mathsf{F}_d$ a diagonal 
distribution function and the inverse ${\sf S}^{-1}$ is defined with respect
to the star product (${\sf S}^{-1}\star{\sf S}={\sf S}\star{\sf
  S}^{-1}=1$, see Sec.~\ref{sec:diagonalization-1} for details).
Note that $\ms F_d$ is real by definition, and we show later that $\ms K_d$ is also real.

There is some phase ambiguity in the diagonalization, which can be
regarded as a gauge freedom since the diagonalization exists at
  every point in phase space.  Physical quantities must be gauge
invariant.  We find that the {\em diagonal} gauge invariant quantities $\ul{\sf O}_d$
take the form 
\begin{align}
  \label{eq:9}
  \ul{\sf O}_d(q)\equiv{\sf O}_d[\ul{\smash{q}}],
\end{align}
with ${\sf O}_d = \mathsf{K}_d,\mathsf{F}_d$, and
\begin{align}
  \label{eq:10}
  \ul{\smash{q}}_\alpha&=q_\alpha+\shbar\epsilon_{\alpha\beta}{\sf
                         A}_\beta+O(\shbar^2),
\end{align}
where $\mathsf{A}_\alpha$ is a Berry gauge field:
\begin{align}
  \label{eq:11}
  \mathsf{A}_\alpha & = \textrm{Im}\, \left(
                      \mathsf{\Lambda}_\alpha\right)^{({\rm d})},
                      \nonumber \\
  \mathsf{\Lambda}_\alpha & = \mathsf{S}^{-1} \star
                      \partial_\alpha \mathsf{S}.
\end{align}
Here $({\rm d})$ denotes the diagonal part and Im the imaginary part.
Note that ${\sf A}$ is diagonal so these equations, e.g. Eq.~\eqref{eq:9}, are unambiguous: see Eq.~\eqref{eq:112a} for an explicit expression
for a diagonal matrix function of a diagonal matrix.

Applying this prescription we obtain the gauge invariant kinetic equation to first order in $\shbar$,
\begin{align}
  \label{eq:191a}
  \partial_t\ul{\sf F}_d+\upsilon\partial_t{\sf
  q}_\alpha\partial_\alpha\ul{\sf F}_d=0 + O(\upsilon\shbar^2),
\end{align}
where we work in the collisionless limit of free particles without
short-distance scatterers although scattering effects could be straightforwardly included on the right-hand-side
  of Eq.~\eqref{eq:191a} in the form of a ``collision integral'' $\mc
  I_{\rm coll}$ \cite{mangeolle2022prx,mangeolle2022prb}. Here $\upsilon = \shbar/\hbar$ is a symbolic parameter that counts
  the order of the semi-classical expansion, shorn of any dimension:
  taken literally $\upsilon=1$ since $\shbar=\hbar$.  The total number
  of powers of $\shbar$ and $\upsilon$ in an expression gives the
  semi-classical order of the corresponding term in the kinetic
  equation, and is equal to the number of spatial gradients. 

In Eq.~\eqref{eq:191a} we have defined the object
\begin{equation}
  \label{eq:196}
  \partial_t{\sf q}_\alpha\equiv
  \epsilon_{\alpha\beta}\left(\partial_\beta\ul{\sf K}_d+\shbar\epsilon_{\gamma\sigma}{\sf
    \Omega}_{\beta\gamma}\partial_\sigma\ul{\sf K}_d\right) +O(\shbar^2).
\end{equation}

Mathematically $\partial_t{\sf q}_\alpha$ is just a function of phase
space given by the right hand side of Eq.~\eqref{eq:196}, i.e.\ there
is no need to solve a dynamical equation for ${\sf q}_\alpha(t)$ to
solve the Boltzmann equation.  However, we can identify physically
Eq.~\eqref{eq:196} as the single particle equation of motion in phase
space of a particle with phase space coordinates ${\sf q}$. We also introduced
\begin{align}
  \label{eq:12}
  {\sf \Omega}_{\alpha\beta} & = \partial_\alpha \mathsf{A}_\beta - \partial_\beta \mathsf{A}_\alpha 
\end{align}
which is the Berry curvature in phase space. Note that this quantity is
  band-diagonal by definition and  ${\sf \Omega}_{\alpha\beta}= {\rm
    Im} \left [ \ms \Lambda_\beta \, \overset \star , \,\ms \Lambda_\alpha
  \right ]^{\rm (d)}$, where we defined the star/Moyal bracket as
$[{\sf A}\,\overset{\star},\,{\sf B}]\equiv{\sf A}\star{\sf B}-{\sf
  B}\star{\sf A}$. This result, which is derived systematically using only the
assumptions in Sec.~\ref{sec:basic-formulation}, reproduces the
expected form from the heuristic wave packet theory \cite{sundaram1999}.

By re-expressing physical quantities in the gauge invariant diagonal
variables, we analogously obtain explicitly gauge-invariant
expressions for them.
The simplest is the energy density in phase space,
\begin{align}
  \label{eq:13}
  \mathcal{H}(q) & =\frac{1}{2}{\rm Tr}\Big[\,\mathfrak{J}\,\ul{\sf F}_d\,\Big]+O(\hbar^2),
\end{align}
where 
\begin{align}
  \label{eq:14}
  \mathfrak{J} = 1 + \shbar \,\sOmega_{X_\mu p_\mu} + O(\shbar^2)
\end{align}
is a Jacobian in phase space, see Sec.~\ref{sec:jacobean}.  It has the
physical implication of inducing on the density of states a dependence on Berry curvature \cite{xiao2005berry}.

Similarly, we obtain the current density in phase space
\begin{equation}
  \label{eq:198}
  \mathcal{J}_\alpha(q) =\frac{\upsilon}{2}{\rm
    Tr}\left[\,\mathfrak{J}\,\partial_t {\sf q}_\alpha\, \ul{\sf F}_d+\shbar
    \epsilon_{\alpha\beta}\epsilon_{\gamma\lambda}\partial_\gamma\left(
      \mathsf{M}_{\beta\lambda}\ul{\sf F}_d\right)\right]+O(\upsilon\shbar^2).
\end{equation}

One should keep in mind that $\alpha$ here is a general phase space index, so that this is more general than the usual current density.
Specifically $\mathcal{J}_{X_\mu}(X,p)$ is the contribution of states at momentum $p$ to the energy current density at position $X$,
while $\mathcal{J}_{p_\mu}(X,p)$ gives the analogous contribution to the ``force density''.

In Eq.~\eqref{eq:198}, we introduced
\begin{equation}
  \label{eq:61a}
  \mathsf{M}_{\alpha\beta} =\frac{1}{2}{\rm Im}
  \left\{ \mathsf{\Lambda}_\beta ,\left[\mathsf{\Lambda}_\alpha ,{\sf K}_d\right]
\right\}^{({\rm d})},
\end{equation}
which has the interpretation of an energy magnetization \cite{cooper1997,qin2011energy}.

\subsection{Practical remarks}
\label{sec:practical-remarks}

Note that in Eq.~\eqref{eq:198} we have separated out two
contributions to the current density, in which the second is
manifestly a total derivative (phase space curl).  The curl term gives
zero when integrated over phase space, and most importantly gives zero
net (momentum integrated) flux through a surface in real space for
which the magnetization $\mathsf{M}_{\beta\lambda}$ vanishes on the
boundary (i.e.\ a surface which cuts the entire sample volume) ---
see also Appendix~\ref{sec:expl-separ-into}.
Thus the energy magnetization
does not contribute to the total `transport current', 
and for
transport purposes only the first term (containing
$\partial_t {\sf q}_\alpha$) need be considered.

To use the above results in a specific problem, one needs to specify
the diagonalized band energy ${\sf K}_d$ and the connection $\mathsf{\Lambda}_\alpha$.
These are obtained by solving the diagonalization problem
perturbatively in $\shbar$.  For the energy, there is a simple general
result to first order,
\begin{align}
  \label{eq:211a}
  \ul{\sf K}_d(q)& ={\sf K}_d[\ul{\smash{q}}] \\
  & = {\sf K}_{0,d}-\shbar\frac{\epsilon_{\alpha\beta}}{2}\mathsf{M}_{\alpha\beta}+\shbar\, {\rm
  Re}\left[\left({\sf S}_0^{-1}{\sf K}_1{\sf S}_0\right)^{({\rm d})}\right]+O(\shbar^2), \nonumber
\end{align}
where ${\sf K}_{0,d}$ is a diagonal matrix containing the conventional
eigenvalues of the matrix
$\mathsf{K}_0 = \hbar\mathsf{\Gamma}\mathsf{H}$, and the connection
$\mathsf{\Lambda}_\alpha$ may be approximated at the same accuracy by
the connection obtained from the this eigenvalue problem at zeroth
order, i.e.\
$\mathsf{\Lambda}_\alpha \approx \mathsf{S}_0^{-1} \partial_\alpha
\mathsf{S}_0$, with
$\mathsf{S}_0^{-1} \mathsf{K}_0 \mathsf{S}_0 = {\sf K}_{0,d}$.  The
final term in the square brackets in Eq.~\eqref{eq:211a} is a
correction arising from the semi-classical expansion of $\mathsf{K}$
itself, with ${\sf K} = {\sf K}_0+\shbar{\sf K}_1+O(\shbar^2)$ (see
App.~\ref{sec:diag-ohbar}). We will show that the spectrum of $\ms K$
is symmetric with positive and negative energy eigenstates related by
complex conjugation and momentum reversal (see
Appendix~\ref{sec:symmetry-spectrum}).

It can be useful both for intuition and as a check to consider the
dimensions of the various objects encountered in this treatment.  This
is somewhat problem dependent, owing to different possible choices of
the fundamental $\Phi_a$ fields, which might be position and momentum
densities for phonons, or transverse magnetization densities in a
ferromagnet, for example.  Thus different components might even have
different dimensions.  However, by appropriate rescaling, it is generally
possible to bring them to a form in which all $\Phi_a$ fields have dimension of the square root of density, $[\Phi_a]=L^{-d/2}$.  Then one obtains
\begin{align}
  \label{eq:122}
  & [X_\mu]=L, \quad [k_\mu]=1/L, \quad [p_\mu]=Et/L,\quad [\mathfrak{J}]=1,\nonumber\\
  & [\mathcal{H}]=[{\sf K}]=[{\sf K}_d]=[{\sf F}_d]=E, \nonumber\\
  & [\mathcal{J}_X]=EL/t, \quad [\mathcal{J}_p]=E^2/L, \nonumber \\
  & [{\sf \Lambda}_\alpha]=[{\sf A}_\alpha]=1/[\alpha],\nonumber\\
  &  [{\sf \Omega}_{\alpha\beta}]=1/([\alpha][\beta]),\quad [{\sf M}_{\alpha\beta}]=E/([\alpha][\beta]), \nonumber\\
  & [{\sf H}][{\sf F}]=E,\quad [{\sf H}][{\sf \Gamma}]=1/t,\quad  [{\sf S}]^2= 1/E,
\end{align}
where $E$, $L$ and $t$ stand for energy, length and time,
respectively. Note that in Eq.~\eqref{eq:122}, all quantities from
$\mathfrak{J}$ onward are defined in phase space.  This brings a
factor of $L^d$ to many quantities owing to the Wigner transform,
which contains a spatial integral, for example
$[\hat{\mathsf{H}}]=E/L^d$ is an energy density while $[\mathsf{H}]=E$
is an energy (if $[\Phi_a]=L^{-d/2}$).  
Note that all the quantities in Eq.~\eqref{eq:122} are of order
$O(\shbar^0)$. This is irrespective of our choice of definitions for
$p_\mu$ and ${\sf K}$ for example, which we chose to be momentum
$p_\mu=\hbar k_\mu=O(\shbar^0)$ and energy ${\sf K}=\hbar\,{\sf \Gamma}\star{\sf
  H}=O(\shbar^0)$ as opposed to wavevector and frequency, respectively.

\section{Detailed derivation}
\label{sec:detailed-derivation}

Here we provide the derivation of the results derived in Section~\ref{sec:overview}.

\subsection{Dynamics and continuity equation}
\label{sec:dynamics}

Together, Eqs.~(\ref{eq:1}-\ref{eq:2}) determine the dynamics of the fields according to the Heisenberg equation of motion,
\begin{equation}
  \label{eq:heiseom}
  \partial_t\Phi_a=-\frac{i}{\hbar}[\Phi_a,{\rm H}] = \frac{1}{\hbar}\left(\hat{\mathsf{K}} \otimes \Phi\right)_a,
\end{equation}
where convolution $\otimes$ is defined as
\begin{equation}
  \label{eq:15}
  (\hat{\sf O}_1\otimes\hat{\sf O}_2)_{ab}(r_1,r_2)\equiv\int_r \sum_{c}\hat{\sf O}_{1,ac}(r_1,r) \hat{\sf O}_{2,cb}(r,r_2),
\end{equation}
and the dynamical matrix is
\begin{equation}
  \label{eq:16}
  \hat{\sf K}\equiv\hbar\,\hat{\sf \Gamma}\otimes\hat{\sf H}.
\end{equation}
In turn, this implies that
\begin{equation}
  \label{eq:17}
  \partial_t\hat{\sf F}=\frac{-i}{\hbar}(\hat{\sf K}\otimes\hat{\sf F}-\hat{\sf
  F}\otimes\hat{\sf K}^\dagger),
\end{equation}
where $(\hat{\sf K}^\dagger)_{ab}(r_1,r_2)\equiv\left(\hat{\sf
    K}_{ba}(r_2,r_1)\right)^*$.
In the same notation the total energy is
\begin{equation}
  \label{eq:18}
  H\equiv\langle{\rm H}\rangle=\frac{1}{2}\int_{r}{\rm Tr}[(\hat{\sf
    H}\otimes\hat{\sf F})(r,r)],
\end{equation}
where ${\rm Tr}$ is the usual matrix trace, i.e.\ ${\rm Tr}[\hat{\sf
  O}]=\sum_a\hat{\sf O}_{aa}$.

Now we pass to the Wigner transform, and use the fact that  the Wigner
transform of a convolution is the star (or Moyal)
product of the Wigner transforms, i.e.\ $[\hat{f}_1\otimes \hat{f}_2]^W(X,p)=f_1(X,p)\star
f_2(X,p)$ (see Appendix~\ref{sec:moyal-product}). This leads directly to Eq.~\eqref{eq:5} and Eq.~\eqref{eq:6}.  

In this way, the total energy can also be written as
  \begin{equation}
    \label{eq:19}
    \langle{\rm H}\rangle\equiv H=\frac{1}{2}\int_{X,p}{\rm Tr}[{\sf H}\star{\sf F}],
  \end{equation}
where ${\rm Tr}[{\sf O}]=\sum_a{\sf O}_{aa}$ for any matrix ${\sf O}$ is the regular
matrix trace.  

Conservation of energy is ensured by Heisenberg evolution, Eq.~\eqref{eq:heiseom}, which is unitary, and when
the Hamiltonian is local, this gives rise to a continuity equation.
To obtain it we first define from Eq.~\eqref{eq:19} a phase space energy density
\begin{equation}
    \label{eq:20}
    \mathcal{H}(X,p)=\frac{1}{4}{\rm Tr}\left({\sf H}\star{\sf
        F}+{\sf F}\star{\sf H}\right)\equiv\frac{1}{4}{\rm Tr}\left\{{\sf H}\,\overset \star ,\, {\sf
        F}\right\}.
\end{equation}
Here we symmetrized by hand the argument of the trace, so that the
energy density is real, $\mathcal{H}=\mathcal{H}^*$ (note
that the choice of {\em local} energy density is not unique
\cite{kapustin2020, hardy1963energy}).

Taking the time derivative of
the energy density, we find
\begin{equation}
  \label{eq:21}
  \partial_t\mathcal{H}=\frac{1}{2\hbar}{\rm Im}{\rm Tr}\left([{\sf
    K}\,\overset{\star},\,{\sf F}\star{\sf H}]\right) .
\end{equation}
Very
generally, the phase space integral of the trace of a Moyal bracket
vanishes (see Appendix~\ref{sec:moyal-product}).  Hence, the right-hand-side of
Eq.~\eqref{eq:21} can can be written as a {\em phase-space} divergence,
i.e.\ there exists $\boldsymbol{\mathcal{J}}(X,p)$ such that, in the
collisionless limit,
  \begin{equation}
    \label{eq:22}
    \partial_t\mathcal{H}+\partial_\alpha\mathcal{J}_\alpha=0.
  \end{equation}
Up to this point, all the above formulas are {\em exact}.
  
Expanding the ``center''
  of the Moyal bracket (i.e.\ ``$\overset{\star},$'') in $\shbar$ in Eq.~\eqref{eq:21}, we find
  \begin{equation}
    \label{eq:23}    \mathcal{J}_\alpha(X,p)=\frac{\upsilon}{2}\epsilon_{\alpha\beta}{\rm
      Re}\, {\rm
      Tr}\left(\partial_\beta{\sf K}({\sf F}\star{\sf H})\right)+O(\upsilon\shbar^2),
  \end{equation}
where ${\rm Re}$ is the real part.  We note that the local phase-space
current density
 $\boldsymbol{\mathcal{J}}$ depends on the precise choice of local energy
 density and enjoys an additional freedom in the form of an additive
 phase-space curl, $\mathcal{J}_\alpha\rightarrow
   \mathcal{J}_\alpha+\partial_\beta\mathcal{M}_{\alpha\beta}$ for any differentiable phase-space function
   $\mathcal{M}_{\alpha\beta}$ odd under $\alpha\leftrightarrow\beta$, 
provided that the curl is {\em gauge invariant}. 
 
 To summarize, the energy density is defined by Eq.~\eqref{eq:20}, and the current density by
 Eq.~\eqref{eq:23}. Now by expressing these quantities in
 the diagonal band basis, we will eventually arrive at
 Eq.~\eqref{eq:13} and Eq.~\eqref{eq:198}.  

 \subsection{Diagonalization}
\label{sec:diagonalization-1}

The effects of curvature and geometry arise from the projection into
the manifold of a band \cite{vanderbilt2018book}. 
Formally the projection is carried out in the phase
space representation by diagonalizing the dynamical matrix using a
similarity transformation $\mathsf{S}$, introduced in Sec.~\ref{sec:semicl-solut}
\begin{align}
  \label{eq:24}
  \mathsf{S}^{-1}\star \mathsf{K} \star \mathsf{S} & = \mathsf{K}_d,
\end{align}
where $\mathsf{K}_d$ is a diagonal matrix (whose entries are {\em
  real}, as shown in Appendix~\ref{sec:form-relat-herm}).  

Because $\mathsf{K}$ is not hermitian, $\mathsf{S}$ cannot be chosen star-unitary, but instead can be chosen to satisfy (see Appendix~\ref{sec:form-relat-herm})
\begin{align}
  \label{eq:26}
  \mathsf{S}^\dagger \star \mathsf{H} \star \mathsf{S} & = 1,
\end{align}
where $1$ is the identity matrix.  Eq.~\eqref{eq:26} is the analog of
the star-unitarity condition for a non-unitary matrix. Note that we define the inverse here in the
sense of the star product, i.e.\ $\mathsf{S}^{-1} \star \mathsf{S} =
\mathsf{S}\star \mathsf{S}^{-1} = 1$, as mentioned in Sec.~\ref{sec:semicl-solut}.  

Eqs.~(\ref{eq:24},\ref{eq:26}) are somewhat formal, due to the non-trivial nature of the star product,
but can be solved order by order in $\shbar$, starting from a zeroth order solution which is a standard matrix diagonalization problem,
and which is guaranteed to exist by the properties of $\mathsf{K}$.

The similarity transformation $\mathsf{S}$ is therefore determined (up
to a gauge freedom we will return to below) by $\mathsf{K}$, and once
it is found, we can, as mentioned in Sec.~\ref{sec:semicl-solut}, define the transformed  density matrix
$\mathsf{F}_d$ as well via
\begin{align}
  \label{eq:27}
  \mathsf{F}  = \mathsf{S} \star \mathsf{F}_d \star \mathsf{S}^{\dagger}.
\end{align}
Note that because $\mathsf{S}$ is not unitary but instead satisfies
Eq.~\eqref{eq:26}, $\mathsf{F}_d$ differs dimensionally from
$\mathsf{F}$ by a factor of energy.  In this ``diagonal frame,''
and in the collisionless limit, one has 
 \begin{equation}
   \label{eq:28}
   \partial_t{\sf F}_d+\frac{i}{\hbar}[{\sf K}_d\,\overset{\star},\,{\sf F}_d]=0.
 \end{equation}
 This equation, which is exact, allows a diagonal and time-independent solution for
 $\mathsf{F}_d$.  This is the standard form of the distribution
 function at and near equilibrium in the following sense.
 There is nothing {\em prohibiting} off-diagonal terms in
 $\mathsf{F}_d$, but such solutions necessarily oscillate on the scale
 of the differences between elements in $\mathsf{K}_d/\hbar$, i.e.\ the mode
 frequencies.  This means that slowly time-varying solutions are
 predominantly diagonal.  Moreover, when weak scattering is included,
 such that the associated relaxation time $\tau$ is longer than these
 oscillation periods, the effects of such oscillations largely average
 out.  In any case, in the present formulation without scattering,
 there is no mixing of the diagonal and off-diagonal parts of
 $\mathsf{F}_d$, so we can focus on the former consistently.  

 \subsection{Gauge transformations}
 \label{sec:gauge-transf}

 \subsubsection{General}
\label{sec:general}

As mentioned above, Eqs.~(\ref{eq:24},\ref{eq:26}) leave some freedom in the choice of $\mathsf{S}$.
In particular, if $\mathsf{S}$ satisfies these equations in the sense that it produces a diagonal $\mathsf{K}_d$, then so too does 
\begin{equation}
  \label{eq:29}
  {\sf S}' =  {\sf S}\star{\sf \Theta},
\end{equation}
produce a diagonal ${\sf K}_d'$, where ${\sf \Theta}$
is a diagonal (star-) unitary matrix,
\begin{align}
  \label{eq:30}
  \mathsf{\Theta}^\dagger \star \mathsf{\Theta} = \mathsf{\Theta}\star \mathsf{\Theta}^\dagger = 1.
\end{align}
Note that the diagonalized matrix $\mathsf{K}'_d$ is not generally equal to $\mathsf{K}_d$.  Rather, if we view the change from $\mathsf{S}$ to $\mathsf{S}'$ as a transformation,
\begin{align}
  \label{eq:31}
  \mathsf{S} \rightarrow \mathsf{S}' =  {\sf S}\star{\sf \Theta},
\end{align}
then
\begin{align}
  \label{eq:32}
  \mathsf{K}_d \rightarrow \mathsf{K}'_d = \mathsf{\Theta}^\dagger \star \mathsf{K}_d \star \mathsf{\Theta},
\end{align}
which is manifestly also diagonal and hermitian.  The same transformation law holds for the density matrix,
\begin{align}
  \label{eq:33}
  \mathsf{F}_d \rightarrow \mathsf{\Theta}^\dagger \star \mathsf{F}_d \star \mathsf{\Theta}.
\end{align}
Physical quantities defined in terms of $\mathsf{H},\mathsf{\Gamma},\mathsf{F}$ must, by construction, be invariant under these gauge transformations.

We will now show, perturbatively in $\shbar$, that such gauge transformations can be interpreted as coordinate changes in phase space.

\subsubsection{First order in $\hbar$}
\label{sec:first-order-hbar}

We can express explicitly
\begin{align}
  \label{eq:34}
  \mathsf{\Theta} & = \textrm{diag} \, \left( e^{i\theta_1},\cdots, e^{i\theta_{2N}}\right),
\end{align}
where the condition of unitarity, Eq.~\eqref{eq:30} perturbatively reduces to
\begin{align}
  \label{eq:35}
  {\rm Im} \, \theta_a = O(\shbar^2).  
\end{align}
Hence to $O(\shbar)$, we can use Eq.~\eqref{eq:34} with real-valued
$\theta_a(q) = \theta_a(X,p)$, which in general are arbitary functions on phase
space.

Using the explicit form of $\mathsf{\Theta}$ in Eq.~\eqref{eq:34}, Eqs.~(\ref{eq:32},\ref{eq:33}) reduce to 
\begin{align}
  \label{eq:36}
  {\sf O}_d(q) &\rightarrow {\sf O}'_d(q) = {\sf
  O}_d(q) -\shbar\epsilon_{\alpha\beta}\partial_\beta\stheta\partial_\alpha{\sf
             O}_d(q)+O(\shbar^2), \nonumber \\
  & = {\sf O}_d[q'] + O(\shbar^2),
\end{align}
for $\mathsf{O}_d = \mathsf{K}_d,\mathsf{F}_d$, where $\stheta = \textrm{diag} \, (\theta_1,\cdots,\theta_{2N})$, and
\begin{align}
  \label{eq:37}
  q'_\alpha & = q_\alpha - \shbar \epsilon_{\alpha\beta} \partial_\beta \stheta.
\end{align}
In Eq.~\eqref{eq:36} and Eq.~\eqref{eq:37}, we have noticed that the
change of ${\sf O}_d$ at order $\shbar$ under a gauge transformation is proportional to
a derivative of ${\sf O}_d$, and hence can be absorbed into an
equivalent shift of its argument.

Note that $q'_\alpha$ in Eq.~\eqref{eq:37} contains $\stheta$ which is a diagonal
matrix, i.e.\ a band-dependent quantity, and so Eq.~\eqref{eq:36}
defines a matrix function of a matrix.  This is not ambiguous because both
the matrix and the matrix argument are diagonal, and so can simply be evaluated for each
band.  More explicitly,  the diagonal matrix function $\mathsf{O}$ of
a diagonal matrix argument $\mathsf{C}$ is defined as
\begin{equation}
  \label{eq:112a}
  {\sf
  O}[{\sf C}]\equiv{\rm
diag}\Big(\big\{{\sf O}_{aa}({\sf C}_{aa})\big\}_a\Big).
\end{equation}

\subsection{Gauge connection and gauge invariant quantities}
\label{sec:gauge-conn-gauge}

From Eq.~\eqref{eq:36} it is evident that $\mathsf{O}_d = \ms F_d, \ms K_d$ are not gauge invariant.
This ultimately reflects the presence of Berry phases.  In particular, we see that the effect of a gauge transformation
is to shift the ``canonical'' variable $q_\alpha$ within the argument of ${\sf O}_d$ by a quantity proportional to $\partial_\beta \stheta$.
This motivates a procedure to restore gauge invariance by modifying the argument of $\mathsf{O}_d$ to compensate for this shift.

This compensation is obtained from the Berry gauge field introduced in
Eq.~\eqref{eq:11}.  Under the gauge transformation
Eq.~\eqref{eq:31}, to zeroth order
\begin{equation}
  \label{eq:38}
  {\sf A}_\alpha\rightarrow{\sf A}_\alpha+\partial_\alpha\stheta+O(\shbar).
\end{equation}
We will see that  $\mathsf{A}_\alpha$ as well as
$\mathsf{\Lambda}_\alpha$ emerge naturally in calculations of physical
quantities.  Comparing Eq.~\eqref{eq:38} and Eq.~\eqref{eq:37} shows
how to compensate the transformation of $\mathsf{O}_d$ by changing its argument.

By combining Eqs.~(\ref{eq:9},\ref{eq:10},~\ref{eq:36},\ref{eq:38}), we can see that $\ul{\sf O}_d(q)$ is gauge invariant to first order,
\begin{align}
  \label{eq:39}
  \ul{\sf O}'_d(q) & = \ul{\sf O}_d(q) + O(\shbar^2).
\end{align}
It is expected that physical quantities can be expressed as manifestly gauge invariant expressions when written in terms of such functions, i.e.\ $\ul{\mathsf{K}}_d, \ul{\mathsf{F}}_d$.

\subsection{Gauge invariant kinetic equation}
\label{sec:gauge-invar-kinet}

To derive the gauge-invariant band-diagonal kinetic equation, we semiclassically expand the Moyal
bracket in
Eq.~\eqref{eq:28}:
\begin{align}
  \label{eq:40}
  \partial_t \mathsf{F}_d & = \upsilon\epsilon_{\alpha\beta} \partial_\alpha
                            {\sf K}_d \partial_\beta \mathsf{F}_d + O(\upsilon\shbar^2).
\end{align}

Next we
take the {\em time} derivative of Eq.~\eqref{eq:9} with
$\mathsf{O}_d=\mathsf{F}_d$, which gives
\begin{align}
  \label{eq:41}
  \partial_t \underline{\mathsf{F}}_d(q) & = \partial_t \mathsf{F}_d[\underline{\smash{q}}],
\end{align}
because $\mathsf{A}_\beta$ is time independent by assumption since
we did not include any time-dependence in ${\rm H}$.  Hence
from Eq.~\eqref{eq:40}
\begin{align}
  \label{eq:42}
  \partial_t \underline{\mathsf{F}}_d(q) & = \upsilon\epsilon_{\alpha\beta}
                                          \left. \partial_\alpha
                                           {\sf K}_d(q) \partial_\beta
                                           \mathsf{F}_d(q)
                                           \right|_{q\rightarrow \underline{\smash{q}}}+O(\upsilon\shbar^2).
\end{align}

Now we need to convert the functions on the right hand side above to
underlined quantities.  To
do so, we take the {\em phase space} derivative of Eq.~\eqref{eq:9},
\begin{align}
  \label{eq:43}
  \partial_\alpha \underline{\mathsf{O}}_d(q) & = \partial_\alpha
                                             \mathsf{O}_d[\ul{\smash{q}}]
                                             + \shbar
                                             \epsilon_{\gamma\lambda}
                                             \partial_\gamma \mathsf{O}_d\,
                                             \partial_\alpha
                                             \mathsf{A}_\lambda+O(\shbar^2),
                                             \nonumber \\
  \Rightarrow \partial_\alpha
                                             \mathsf{O}_d[\ul{\smash{q}}]
                                           & = \partial_\alpha
                                             \underline{\mathsf{O}}_d(q)
                                             - \shbar
                                             \epsilon_{\gamma\lambda}
                                             \partial_\gamma
                                             \underline{\mathsf{O}}_d \,
                                             \partial_\alpha
                                             \mathsf{A}_\lambda +
                                             O(\shbar^2).  
\end{align}
Using Eq.~\eqref{eq:43} for both ${\sf K}_d$ and $\mathsf{F}_d$ in the
right hand side of Eq.~\eqref{eq:42}, we arrange terms to obtain
\begin{align}
  \label{eq:44}
  \partial_t \underline{\mathsf{F}}_d & = \upsilon\bigg[\epsilon_{\alpha\beta} \partial_\alpha
                            \underline{\sf K}_d \partial_\beta
                            \mathsf{F}_d - \shbar
                            \epsilon_{\alpha\beta}
                            \epsilon_{\gamma\lambda} \partial_\gamma
                            \underline{\sf K}_d \partial_\beta
                            \underline{\mathsf{F}}_d \partial_\alpha
                            \mathsf{A}_\lambda \nonumber \\
  & \quad- \shbar \epsilon_{\alpha\beta} \epsilon_{\gamma\lambda}
    \partial_\alpha\underline{\sf K}_d \partial_\gamma
    \underline{\mathsf{F}}_d \partial_\beta \mathsf{A}_\lambda + O(\shbar^2)\bigg].
\end{align}
Now relabeling dummy indices in the last term above according to $\alpha\rightarrow \gamma$, $\beta
\rightarrow \lambda$, $\gamma\rightarrow \beta$, $\lambda \rightarrow
\alpha$, one obtains Eq.~\eqref{eq:191a} with $\partial_t
{\sf q}_\alpha$
as in Eq.~\eqref{eq:196}. 

\subsection{Gauge invariant energy and current density}
\label{sec:gauge-invar-energy}

Now we proceed to derive diagonal and gauge invariant expressions for the energy density, Eq.~\eqref{eq:20} and
energy current density, Eq.~\eqref{eq:23}.

\subsubsection{Energy density}
\label{sec:energy-density}

Let us begin with the energy density.  Using properties of the trace and others recapped in
Appendix~\ref{sec:moyal-product}, we write Eq.~\eqref{eq:20} as
\begin{align}
  \label{eq:45}
  \mathcal{H} & = \frac12 \textrm{Re}\,\textrm{Tr}\, \left(
                \mathsf{F}\star\mathsf{H}\right) \nonumber \\
  & = \frac12 \textrm{Re}\,\textrm{Tr}\, \left(\mathsf{S} \star
    \mathsf{F}_d \star \mathsf{S}^{-1}\right),
\end{align}
using Eqs.~(\ref{eq:24},\ref{eq:26},\ref{eq:27}).

To simplify this, we use the identity, valid for any operator ${\sf O}$,
\begin{align}
  \label{eq:46}
  \textrm{Tr}\, \left(\mathsf{S} \star
    \mathsf{O} \star \mathsf{S}^{-1}\right) & = \textrm{Tr}\,
                                              \mathsf{O} + i
                                              \shbar
                                              \epsilon_{\alpha\beta} \partial_\beta
                                              \textrm{Tr} \, \left(
                                              \mathsf{\Lambda}_\alpha
                                              \mathsf{O}\right) + O(\shbar^2),
\end{align}
which is derived straightforwardly by the expansion of the star
product.  Taking $\mathsf{O}=\mathsf{F}_d$ and applying this to
Eq.~\eqref{eq:45}, we then obtain
\begin{align}
\label{eq:47}
  \mathcal{H}&=\frac{1}{2}\left({\rm Tr}{\sf F}_d-\shbar\epsilon_{\alpha\beta}\partial_\beta{\rm
      Tr}\left[{\sf A}_\alpha{\sf
    F}_d\right]\right)+O(\shbar^2)\\
 & =\frac{1}{2}{\rm Tr}\Big[\left(1+\frac{\shbar}{2}\epsilon_{\gamma\lambda}{\sf \Omega}_{\gamma\lambda}\right)\left({\sf F}_d+\shbar\epsilon_{\alpha\beta}{\sf
    A}_\beta\partial_\alpha{\sf F}_d\right)\Big]+O(\shbar^2),\nonumber
\end{align}
where we defined (the band-diagonal quantity) ${\sf \Omega}_{\alpha\beta}$ in Eq.~\eqref{eq:12}. This expression is diagonalized and gauge-invariant but 
the latter gauge invariance is not explicit.  To make it so, we use the definition in
Eq.~\eqref{eq:9} and Taylor expand in $\shbar$ {\em at fixed} $q_\alpha$.
This gives
\begin{align}
  \label{eq:48}
  \underline{\mathsf{O}}_d(q) & =
 \mathsf{O}_d(q) +
    \shbar\epsilon_{\alpha\beta}\mathsf{A}_\beta \partial_\alpha
    \mathsf{O}_d(q) + O(\shbar^2).                   
\end{align}
Applying Eq.~\eqref{eq:48} with $\mathsf{O}_d=\mathsf{F}_d$ immediately gives
Eqs.~(\ref{eq:13},\ref{eq:14}) in Sec.~\ref{sec:overview}.
The fact that $\mf J$, defined in Eq.~\eqref{eq:14}, identifies with the jacobian of the
$q \mapsto \underline{\smash q}$ transformation is shown in App.~\ref{sec:jacobean}.

\subsubsection{Energy current}
\label{sec:energy-current}

Now we turn to the energy current, written in Eq.~\eqref{eq:23}.   
It
contains two factors inside the trace, $\partial_\beta {\sf K}$, and
$\mathsf{F}\star\mathsf{H}$, which we want to write in diagonal form.  To express the former, we note the useful identity
\begin{align}
  \label{eq:50}
  \partial_\alpha \left( \mathsf{S}\star \mathsf{O}\star
  \mathsf{S}^{-1}\right) & = \mathsf{S}\star \left( \partial_\alpha
                           \mathsf{O} + \left[ {\sf \Lambda}_\alpha
                           \overset{\star}, \,\mathsf{O}\right]\right)
                           \star \mathsf{S}^{-1}.
\end{align}
Applying Eq.~\eqref{eq:50} with $\mathsf{O} = {\sf K}_d =\mathsf{S}^{-1}\star
{\sf K} \star \mathsf{S}$ gives
\begin{equation}
\label{eq:51}
  \partial_\alpha{\sf K} ={\sf S}\star {\sf Q}_\alpha\star{\sf S}^{-1},
\end{equation}
where we defined for convenience
\begin{equation}
  \label{eq:52}
  {\sf Q}_\alpha\equiv \partial_\alpha{\sf K}_d+\left[{\sf
      S}^{-1}\star\partial_\alpha{\sf S}\,\overset{\star},\,{\sf K}_d\right].
\end{equation}
The second multiplicative factor in the trace in Eq.~\eqref{eq:23} is, using Eqs.~(\ref{eq:26},\ref{eq:27}),
\begin{align}
  \label{eq:53}
  \mathsf{F}\star\mathsf{H} & = \mathsf{S} \star \mathsf{F}_d \star \mathsf{S}^{-1}.
\end{align}
Using Eq.~\eqref{eq:51} and Eq.~\eqref{eq:53}, one can convert
Eq.~\eqref{eq:23} to diagonal form using manipulations similar to
those for the energy density, though with more involved algebra.  The
details are left to Appendix~\ref{sec:diagonalization}.  One obtains in this way
\begin{equation}
  \label{eq:565} \mathcal{J}_\alpha=\mathcal{J}_\alpha^{(1)}+\mathcal{J}_\alpha^{(2)},
\end{equation}
in which each  term is separately gauge invariant:
\begin{subequations}
  \begin{align}
\mathcal{J}_\alpha^{(1)}&=\frac{\upsilon}{2}\epsilon_{\alpha\beta}{\rm
                            Tr}\Big[\big(1+\frac{\shbar}{2}\epsilon_{\gamma\lambda}{\sf
                                           \Omega}_{\gamma\lambda}\big)\nonumber \\
  & \quad\times\big(\partial_\beta({\sf K}_d+\shbar\epsilon_{\rho\sigma}{\sf
                            A}_\sigma\partial_\rho{\sf K}_d)+\shbar\epsilon_{\rho\sigma}{\sf
                            \Omega}_{\beta\rho}\partial_\sigma{\sf K}_d\big)\nonumber
  \\
  & \quad\times\big({\sf
                            F}_d+\shbar\epsilon_{\xi\eta}{\sf
                            A}_\eta\partial_\xi{\sf F}_d\big)\Big]+O(\upsilon\shbar^2),   \label{eq:66a} \\
  \label{eq:66b}
  \mathcal{J}_\alpha^{(2)}&=\frac{\upsilon\shbar}{2}\epsilon_{\alpha\beta}\epsilon_{\gamma\lambda}\partial_\gamma{\rm
                            Tr}\big[\mathsf{M}_{\beta\lambda}{\sf
                            F}_d\big]+O(\upsilon\shbar^2),
  \end{align}
\end{subequations}
where $\mathsf{M}_{\alpha\beta}$ was defined in Eq.~\eqref{eq:61a}. Now we can again use Eq.~\eqref{eq:48} to identify gauge-invariant
functions $\underline{\sf K}_d$ and $\underline{\mathsf{F}}_d$,
and we directly thereby obtain Eq.~\eqref{eq:198} in Sec.~\ref{sec:overview}.

Let us discuss the physical meaning and relevance of
$\mathcal{J}_\alpha^{(2)}$.  This term satisfies the continuity equation \emph{by definition}, as it is a phase space ``curl''.
Such terms appear to be ambiguous from the introduction of the heat current, which occurred in passing from Eq.~\eqref{eq:21} to Eq.~\eqref{eq:23}.
That ambiguity arises because the quantity which enters the continuity equation is only the divergence of the phase space current,
so that a shift $\mathcal{J}_\alpha\rightarrow \mathcal{J}_\alpha+\partial_\beta\mathcal{M}_{\alpha\beta}$ with antisymmetric $\mc M$, leaves the continuity equation unchanged.

The above derivation certainly provides an unambiguous result for $\mathcal{J}_\alpha^{(2)}$ given the definition in Eq.~\eqref{eq:23}.
However, we would like to ask whether the result in Eq.~\eqref{eq:66b} can be considered to resolve the intrinsic ambiguity.
To resolve it, one must clearly impose some other conditions.
Related conditions are discussed for the energy magnetization itself in Ref.~\cite{qin2011energy}.
For the energy magnetization current, we are not sure of the full set of conditions that should be imposed, but we do note that the above definition satisfies many reasonable ones:
\begin{enumerate}
\item The quantity ${\sf M}_{\alpha\beta}$ itself is a gauge-invariant
band-diagonal quantity which can be identified as an energy
magnetization. Then, $\mathcal{J}^{(2)}_\alpha$ is naturally the
gauge-invariant divergenceless current which can be built from it, and
it indeed directly appears when extracting the divergence from the
time-derivative of the local energy density
$\partial_t\mathcal{H}$.
\item The magnetization contribution,  $\mathcal{J}_\alpha^{(2)}$ vanishes in equilibrium when the system is locally uniform.
  We may define equilibrium to be such
that the energy density ${\sf F}_d$ depends only on the band energy
${\sf K}_d$, i.e.\ ${\sf F}_d(q)=f[{\sf K}_d(q)]$, then
 $\partial_\gamma{\sf F}_d=\partial_\gamma {\sf K}_d f'[{\sf K}_d(q)]$.  Then
$\mathcal{J}^{(2)}_\alpha$ vanishes if $\partial_X{\sf K}_d=0$ and $\partial_X{\sf S}=0$.  It does not, however, a priori vanish otherwise, even in equilibrium. A
local nonvanishing current in equilibrium is in fact how the
magnetization current has been defined in the literature \cite{cooper1997}.
\item One might expect that in equilibrium, even the local thermal current $J_{X_\mu} = \int_p ( \mc J_{X_\mu}^{(1)} + \mc J_{X_\mu}^{(2)} )$, vanishes at zero temperature.
  This would seem natural since the entropy vanishes at zero temperature.
  We are able to show that this is the case in the special but important case of a $(p,X)$-separable problem such as we describe below and in Appendix \ref{sec:case-grav-field}.
\item As noted in Refs.~\cite{qin2011energy,qin2012}, the momentum
  integral defining $J_{X_\mu}$ is rid of a UV divergence when
  evaluated in a continuum theory, in the $(p,X)$-separable case.
  This is not the case for the integrands $\mc J_{X_\mu}^{(1)} $ and
  $\mc J_{X_\mu}^{(2)} $ considered separately.  Thus
  Eq.~\eqref{eq:66b} is ``physical'' in the sense that it provides a
  compensating term to an otherwise unphysical divergence in
  Eq.~\eqref{eq:66a}.
\end{enumerate}

\section{Application: thermal Hall current in a finite system of chiral bosons}
\label{sec:applications}

We now show how our QKE formalism can be applied to any general system
with a boundary but homogeneous away from the boundary, and further
specialize to a system inhomogeneous linear elasticity theory with a
phonon Hall viscosity term.

\subsection{Uniform finite system with a boundary: formal result for thermal conductivity}
\label{sec:unif-finite-syst}

One natural application of our formalism is to a finite system, in which the presence
of a boundary is included via 
variations of the Hamiltonian in space.
Specifically, we assume that the (2d) system is constrained within the interval
$I_x = \left \{ x \in \left [ - \frac L 2, +\frac L 2 \right ] \right \} $,
and we consider a Hamiltonian of the form ${\sf H}(X,p)=g(X){\sf H}^{\rm h}(p)$,
with $g$ a c-number function equal to 1 in the $I_x$ interval and
decaying over distances $O(L^0)$ to 0 outside,
mimicking the presence/absence of physical degrees of freedom in these regions.
This is just a specific instance of the more general problem of a $(X,p)$-separable theory,
which we consider in detail in Appendix~\ref{sec:case-grav-field}.

There, we show that solving the separable problem amounts to solving
the homogeneous one (i.e.\ $g(X)=1$ everywhere), which we label with
${\rm h}$ superscripts:
\begin{subequations}
  \begin{align}
  \label{eq:24hom}
  (\mathsf{S}^{\rm h})^{-1}\mathsf{K}^{\rm h} \mathsf{S}^{\rm h} & = \mathsf{K}_d ^{\rm h},\\
    \label{eq:26hom}
  (\mathsf{S}^{\rm h})^\dagger\mathsf{H}^{\rm h} \mathsf{S}^{\rm h} & = 1 ,
\end{align}
\end{subequations}
where the $\star$-product was replaced by the standard matrix product in this $X$-independent problem,
and $\ms K^{\rm h} = \hbar\,\ms \Gamma^{\rm h}\ms H^{\rm h}$. This is now a standard linear algebra problem,
yielding in particular the homogeneous energies $\ms K_d^{\rm h}$ and Berry curvatures ${\sf \Omega}^{\rm h}_{p_y p_x}$.

We assume translational invariance along $y$ and consider a temperature gradient $\partial_x T$ along the $x$ direction.
The (``transport'') thermal conductivity relates the latter to the net
energy flux flowing in response---via Fourier's law
$\boldsymbol{J}_{\rm tot}=-\boldsymbol{\kappa}^{\rm tr}\cdot\boldsymbol{\nabla}T$---
which is obtained by integrating the energy current along the $x$ direction and taking the thermodynamic limit:
\begin{align}
  \label{eq:115}
  \kappa_{xy}^{\rm tr} &= \frac {1} {\partial_x T} \lim_{L\rightarrow \infty} \frac 1 L \int_{-\infty}^{+\infty} \text dx \int_p \mc J_{X_y}(X,p).
\end{align}
Note that we denote here $(X_x,X_y)\equiv(x,y)$, and we dropped the $O(\shbar^2)$ correction, since $\kappa_{xy}^{\rm
  tr}$ is by definition a linear response quantity. The explicit form
of both currents from the decomposition Eq.~\eqref{eq:565} is derived in App.~\ref{sec:energy-current-1}.

Because $\mc J_{X_y}^{(2)}$ is a total real-space derivative, it does not contribute to the transport current.
Similarly, the zeroth-order term $g \partial_{p_y} {\sf K}_d^{\rm h}$ in $\mc J_{X_y}^{(1)}$ drops out as it is a total momentum-space derivative.
The only contribution which does not vanish after performing the
momentum integral is
\begin{align}
  \label{eq:116}
\mathcal{J}_{X_y} ^{(\rm eff)}&\equiv - \frac{\shbar }{2} \psi_x {\rm
                                Tr}\Big[ {\sf \Omega}^{\rm h}_{p_y
                                p_x} g{\sf K}_d^{\rm h} f(g {\sf K}_d^{\rm h},T) \Big]+O(\shbar^2),
\end{align}
where $\psi_x\equiv \partial_xg/g$ and $ f(\varepsilon,T) = \varepsilon \left (
  n_{\rm B}(\varepsilon,T)+ \tfrac 1 2 \right )$, with $n_{\rm B}(\varepsilon,T)=1/(e^{\varepsilon/(k_{\rm B}T)}-1)$ the
Bose function ($k_{\rm B}$ is Boltzmann's constant), is the equilibrium energy distribution evaluated at $\varepsilon$ of a
system of bosons at temperature $T$. 
Further decomposing explicitly $\rm Tr = Tr_+ + Tr_-$ where $\rm Tr_+$
(resp.\ $\rm Tr_-$) sums over positive (resp.\ negative) energy bands (see Appendix~\ref{sec:symmetry-spectrum}),
and the fact that ${\sf \Omega}^{\rm h}_{p_y p_x} , {\sf K}_d^{\rm h}$
are odd upon spectrum reflection while $f$ is even (note that
$f(-\varepsilon,T)=f(\varepsilon,T)$ and see Appendix~\ref{sec:symmetry-spectrum}), allows one to replace $\rm Tr \mapsto 2 Tr_+$ in Eq.~\eqref{eq:116}.

Importantly, the gradient of temperature is not introduced
``artificially'' in the form of a gravitational field like in
Luttinger's trick \cite{luttinger1964theory} (in particular it is unrelated to $\psi_x$),
but simply arises from the fact that the function $f$ depends
implicitly on $x$ since the Bose function must be computed at the
\emph{local equilibrium} temperature $T(x)$ \cite{cooper1997}.

Eqs.~(\ref{eq:115},\ref{eq:116}) are our solution to the problem, which is readily amenable to numerical evaluation.
It is also possible to reconcile it with existing literature with just one extra step.
Performing the spatial integration with a few simple tricks to turn it
into an integral over energies $\varepsilon$ (cf.\ Appendix~\ref{sec:expl-spat-integr}),
we arrive at the result
\begin{align}
  \label{eq:117}
   \kappa_{xy} ^{\rm tr} &= - \frac 1 T {\rm Tr}_+ \int_p  \ms \Omega_{p_x p_y}^{\rm h}
                          \int \text d \varepsilon \, \Theta (\varepsilon-
                           {\sf K}_d^{\rm h} ) \,\varepsilon^2 \partial_\varepsilon n_B \left ( \varepsilon, T\right ) ,
\end{align}
where $\Theta$ is the step function and $T$ the mean temperature
within $I_x$.
This formula Eq.~\eqref{eq:117} agrees with prior results \cite{qin2012, matsumoto2014}
which were obtained using the Kubo formalism.

\subsection{Local energy current in a chiral phonon system}
\label{sec:local-energy-current}

We now consider a concrete example.  
The model is that of elasticity with a time-reversal breaking term,
described by the (euclidian) Lagrangian
\begin{equation}
  \label{eq:98}
  \mathcal{L}_{\rm ph}=\frac{\rho}{2} \left(\partial_\tau \bs u \right)^2 + \frac{1}{2} c_{ij\mu\nu}
    \partial_\mu u^i \partial_\nu u^j + i \eta_{ij \mu\nu}
    \partial_\tau u^i \partial^2_{\mu \nu} u^j,
\end{equation}
where $\bs u = ( u^i ) _i$ is the displacement field ($i=x,y,z$), $\rho$ is the mass density,
$c_{ij\mu\nu}$ is the bulk modulus ($j,\mu,\nu=x,y,z$), and $\eta_{ij\mu\nu}$ is the
lattice (or phonon) Hall viscosity.
To remove ambiguity we take $c_{ij\nu\mu}=c_{ij\mu\nu}=c_{ji\mu\nu}$
and $\eta_{ij\nu\mu}=\eta_{ij\mu\nu}=-\eta_{ji\mu\nu}$. 

The Hall viscosity term breaks time reversal symmetry explicitly, and accounts for the chiral nature of bosonic excitations in the system (chiral phonons).
The physical origin of this term has been discussed elsewhere in various contexts
such as quantum Hall systems \cite{barkeshli2012}, magnetic insulators \cite{ye2021}, ionic crystals \cite{saito2019berry}, etc.
He we simply use Eq.~\eqref{eq:98} as our starting point, as an effective theory for chiral phonons.

The full inhomogeneous problem, where all parameters $\rho, c_{ij\mu\nu}, \eta_{ij\mu\nu}$ are arbitrary functions of position,
can be recast into the form of Eqs.~(\ref{eq:1},\ref{eq:2}) and can be solved explicitly
following the procedure of Secs.~\ref{sec:overview}, \ref{sec:detailed-derivation}.
We leave the details of calculations and explicit expressions to Appendix~\ref{sec:example-inhom-elast}.

Here in the main text, we restrict ourselves to the spatial dependence
described in the previous subsection. 
The details are left to Appendix~\ref{sec:case-grav-field}, but it
suffices to then study the homogeneous problem,
with position-independent parameters and matrices $\ms H^{\rm h}, \ms
\Gamma^{\rm h}$ ($\Phi=(u,\Pi/\rho)$, with $\Pi_i$ the canonically
conjugate momentum to $u_i$, see Eq.~\eqref{eq:100})
which in the case of the Lagrangian Eq.~\eqref{eq:98} take the simple
forms 
\begin{align}
  \label{eq:121}
 {\sf \Gamma}^{\rm h}=\frac{i}{\hbar\rho}
  \begin{bmatrix}
    0&1\\
    -1& - \frac{2 \eta_{\mu\nu}}{\hbar^2\rho}  p_\mu p_\nu 
  \end{bmatrix} ,\quad
        {\sf H}^{\rm h}=\begin{bmatrix}
   \frac 1 {\hbar^2}  c_{\mu\nu}p_\mu p_\nu&0\\
      0&\rho
      \end{bmatrix},
\end{align}
where $(c_{\mu\nu})_{ij}\equiv c_{ij\mu\nu}$,
$(\eta_{\mu\nu})_{ij}\equiv \eta_{ij\mu\nu}$, and so each block in
Eq.~\eqref{eq:121} is a $3\times 3$ matrix, and $\ms K^{\rm h}=\hbar\,\ms \Gamma^{\rm h}\ms H^{\rm h}$.
For concreteness, we choose the following expressions for the Hall
viscosity and bulk modulus:
\begin{align}
  \eta_{ij\mu\nu} &= \epsilon^{ijz} \delta_{\mu\nu} \left ( \eta_1 \overline \delta_{\nu z} + \eta_2 \delta_{\nu z} \right ),\nonumber\\
   c_{ij\mu\nu}&= c_1 \delta_{ij}\delta_{\mu\nu}
                + \delta_{i\mu}\delta_{j\nu} \left ( c_2 \overline \delta_{iz} \overline \delta_{jz} + c_3  \delta_{iz}\delta_{jz} \right ) \nonumber\\
                  &+ \delta_{\mu\nu} \left ( c_4  \overline \delta_{\nu z}  \delta_{iz}\delta_{jz} 
                    +  c_5 \delta_{\nu z} \overline \delta_{iz} \overline \delta_{jz} \right ) ,\label{eq:122b}
\end{align}
where $\epsilon^{ijl}$ is the 3d Levi-Civita tensor, $\overline \delta_{ab} \equiv 1 - \delta_{ab}$,
and $\eta_1, \eta_2, c_a, a=1..5$, are real parameters.
This makes $\mathcal{L}_{\rm ph}$ the most general possible Lagrangian
preserving $O(2)$ symmetry around the $\mb{\hat z}$ axis, except that,
for simplicity, we do {\em not} include terms which couple the $u^z$
to the $(u^x,u^y)$ components, even if they preserve $O(2)$ symmetry. 
This allows us to reduce the problem
to be effectively a $4\times4$ problem in matrix space.

We leave details of the solution of the diagonalization problem to the appendix,
and show in Figure~\ref{fig:local-currents} the position-resolved
momentum-integrated energy currents for a given 
profile of $g$ and the parameter values provided in Table~\ref{tab:parameter-values}.

\begin{table}[htbp]
  \centering
  \begin{tabular}{c|c|c|c|c|c|c|c|c|c|c|c}
    \hline\hline
     $T_0$  &  $\delta T$  &  $\rho$  &  $c_1$  &  $c_2$  &  $c_3$  &  $c_4$  &  $c_5$  &  $\eta_1$  &  $\eta_2$ & $\xi$  & $L$ \\   \hline
   ${\mathtt{1.0}}$ & $\mathtt{0.8}$ & $\mathtt{1.0}$ & $\mathtt{1.5}$ &$\mathtt{1.2}$ & $\mathtt{0.8}$ & $\mathtt{0.4}$ & $\mathtt{0.0}$ & $\mathtt{-0.25}$ & $\mathtt{-0.15}$ &$\mathtt{1.0}$&$\mathtt{10}$\\
    \hline\hline
  \end{tabular}
  \caption{Values used 
    for the parameters defined in Eqs.~(\ref{eq:98},\ref{eq:122b}) and in the caption of Figure~\ref{fig:local-currents}.}
  \label{tab:parameter-values}
\end{table}

As is evident in Figure~\ref{fig:local-currents}, the magnetization current $\mathcal{J}_{X_y}^{(2)}(x)$ is not zero in the bulk
(the exaggerated temperature gradient we used is made to render this visible), but its integral over a slice of material vanishes exactly.
Meanwhile, the chiral current $\mathcal{J}_{X_y}^{(1)}(x)$ is exactly
zero in the bulk, but the counterpropagating edge densities yield a
finite integral, i.e.\ a nonzero Hall conductivity. 

\section{Conclusion}
\label{sec:conclusion}

In this paper, we presented a systematic derivation of the quantum kinetic equation in the full phase space
for any quadratic Hamiltonian of bosonic fields.  Translation invariance is not assumed, and boundaries may be included.
The treatment provides a derivation of how the quasi-classical distribution function descends from the full quantum density matrix,
and how the evolution of these quantities is related to single-particle characteristics such as the spectrum,
Berry curvature, and energy magnetization.  To obtain these results, we follow the exact statistical formulation of quantum dynamics using the Moyal product in phase space,
and thereby obtain exact equations that can be expanded systematically order by order in a semiclassical parameter such as the smallness of spatial gradients.

From this procedure, in the leading approximation we recover the full
Boltzmann equation and provide a self-contained and exact derivation
of the intrinsic thermal Hall effect of bosons.  Moreover, the
procedure allows a spatially resolved examination of the associated
thermal currents, treating boundaries and the spatial profile they
generate.  For the intrinsic thermal Hall effect, this method
clarifies the separation of transport and ``magnetization'' currents
transparently, and without the need to introduce any artificial
gravitational field which is required in a Kubo formulation following
Luttinger \cite{luttinger1964theory}.

Aside from pedagogical value, we believe this formalism will be useful
in simplifying and clarifying future calculations.  It allows a direct
application to intrinsically inhomogeneous systems such as skyrmionic
textures in chiral magnets, which will be addressed in a forthcoming
publication.  It also offers a natural extension to higher orders in
the semi-classical expansion, in which deviations from Boltzmann 
transport become evident.  This is important for example in
non-linear transport effects (i.e.\ beyond linear order in applied
electric fields and thermal gradients), since such effects are indeed
higher order in gradients. Existing theory of such
effects reveals a role for geometric properties of bands such as Berry
curvature and quantum metric, but because it is based on Boltzmann
transport which is only formally exact up to \emph{first} order in phase space gradients,
the theory is incomplete and should be revisited.

A natural important future direction is the inclusion of scattering and interactions into the present formalism.
Scattering can be straightforwardly added in an \emph{ad hoc} manner, through a collision integral, such as that provided in 
Refs.~\cite{mangeolle2022prx,mangeolle2022prb}.   It would also be
interesting to derive it rigorously from the Keldysh formalism, which
is a natural extension with some precedents \cite{keldysh1964, prange1964transport, rammer, kamenev2011field}.
Another relevant extension is to time-dependent Hamiltonians, such as
would occur in a sliding texture like a driven spin density wave or
skyrmion lattice, which leads to Berry phase effects in spacetime.
Moreover, the same methods can be applied to fermions, as in
Ref.~\cite{wickles2013} for normal electrons, and extended to
Bogoliubov-de Gennes quasiparticles for the superconducting case.
This potentially provides a new way to address phenomena in
topological superconductors.

\acknowledgements

This project was funded by the European Research Council (ERC) under the European Union's Horizon 2020 research and innovation program (Grant agreement No.\ 853116, acronym TRANSPORT).
L.B.\ was supported by the DOE, Office of Science, Basic Energy Sciences under Award No.\ DE-FG02-08ER46524.  We acknowledge the hospitality of
the KITP, where we carried out part of this project, as one does, funded under NSF
Grant NSF PHY-1748958.

\bibliography{Submission.bib}

\appendix

\section{Properties of the Moyal Product}
\label{sec:moyal-product}

The Moyal, or star, product  is widely used in
the phase space formulation of quantum mechanics, and is particularly
convenient for semi-classical analysis.  It is defined in the main
text, Eq.~\eqref{eq:7}, and can be applied to functions, or,
component-wise, to tensor multiplication.  We review some key useful
properties of the Moyal product here.

Even for scalar functions,
it is non-commutative, reflecting the non-commutativity of operators
in quantum mechanics.  It is, however, associative,
\begin{equation}
  \label{eq:54}
  {\sf O}_1\star{\sf O}_2\star{\sf O}_3=({\sf O}_1\star{\sf O}_2)\star{\sf O}_3={\sf O}_1\star({\sf O}_2\star{\sf O}_3).
\end{equation}
Under hermitian
conjugation, we have
\begin{equation}
  \label{eq:55}
  \left({\sf O}_1\star{\sf O}_2\right)^\dagger={\sf O}_2^\dagger\star{\sf O}_1^\dagger,
\end{equation}
where the hermitian conjugate $\dagger$ is defined in the usual matrix
sense, and the phase space argument is unchanged.

The star product is not cyclic in the trace, but
the combination of the trace and phase space ($q\in\mathbb{R}^{2d}$)
integral is cyclic if ${\rm Tr}\left[{\sf O}_1\star{\sf O}_2\right]$
is integrable:
\begin{equation}
  \label{eq:56}
  \int_{q}{\rm Tr}\left[{\sf O}_1\star{\sf O}_2\right]=\int_{q}{\rm Tr}\left[{\sf O}_2\star{\sf O}_1\right].
\end{equation}
This implies that the phase space integral of a Moyal bracket vanishes
\begin{align}
  \label{eq:57}
   \int_{q}{\rm Tr}\left[{\sf O}_1 \overset{\star}, \, {\sf
  O}_2\right] & = 0.
\end{align}
If the functions ${\sf
  O}_{1,2}$ are differentiable, the chain rule applies, i.e., for $\alpha=1,..,2d$,
\begin{equation}
  \label{eq:58}
  \partial_{q_\alpha}({\sf O}_1\star{\sf O}_2)=\partial_{q_\alpha}{\sf
    O}_1\star{\sf O}_2+{\sf O}_1\star\partial_{q_\alpha}{\sf O}_2.
\end{equation}
A matrix function ${\sf O}$ is star-invertible if there exists ${\sf
  O}^{-1}$ such that
\begin{equation}
  \label{eq:59}
  {\sf O}\star{\sf O}^{-1}={\sf O}^{-1}\star{\sf O}=1.
\end{equation}
Note that, as for ordinary matrix multiplication, the left and right
star-inverses are identical.  This follows from viewing the
convolution as a matrix product in the combinated coordinate-index
space, and then using the fact that the Wigner transform of  a
convolution is the star product of Wigner transforms.

If ${\sf O}_1$ and ${\sf O}_2$ are star-invertible,
\begin{equation}
  \label{eq:60}
  ({\sf O}_1\star{\sf O}_2)^{-1}={\sf O}_2^{-1}\star{\sf O}_1^{-1}
\end{equation}
Up to first order in $\shbar$, we have:
\begin{equation}
  \label{eq:61}
  {\sf O}_1\star{\sf O}_2={\sf O}_1{\sf
    O}_2+i\shbar\frac{\epsilon_{\alpha\beta}}{2}\partial_\alpha{\sf
    O}_1\partial_\beta{\sf O}_2+O(\shbar^2).
\end{equation}
The zeroth order term is just standard matrix
multiplication while the first order one is related to the Poisson bracket: $\epsilon_{\alpha\beta}\partial_\alpha{\sf
  O}_1\partial_\beta{\sf O}_2=\partial_{X_\mu}{\sf
  O}_1\partial_{p_\mu}{\sf O}_2-\partial_{p_\mu}{\sf
  O}_1\partial_{X_\mu}{\sf O}_2\equiv\{{\sf O}_1,{\sf O}_2\}_{\rm p.b.}$. 
Eq.~\eqref{eq:61} can be rewritten, up to first
order:
\begin{equation}
  \label{eq:62}
  {\sf O}_1{\sf O}_2={\sf O}_1\star{\sf
    O}_2-i\frac{\shbar}{2}\epsilon_{\alpha\beta}\partial_\alpha{\sf
    O}_1\partial_\beta{\sf O}_2+O(\shbar^2).
\end{equation}
From the above, we can find the ordinary product of a matrix with its star-inverse
\begin{align}
\label{eq:63}
  {\sf O}^{-1}{\sf
  O}&=1-i\frac{\shbar}{2}\epsilon_{\alpha\beta}\partial_\alpha{\sf
      O}^{-1}\partial_\beta{\sf O}+O(\shbar^2),
\end{align}
which notably is not the identity.

\section{Diagonalization of ${\sf K}$}
\label{sec:diagonalization-k}

\subsection{Formal relation to a hermitian problem}
\label{sec:form-relat-herm}

Here we show that the star-diagonalization of $\mathsf{K}$ is related to a
hermitian star-eigenvalue problem, provided the original physical
problem is well-defined.  In particular, this requires that ${\sf H}$
is positive definite.  
The latter implies that there exists a positive definite
matrix ${\sf H}^{1/2}$ such that
\begin{equation}
  \label{eq:64}
  {\sf H}^{1/2}\star{\sf H}^{1/2}={\sf H}.
\end{equation}
We define the matrix $\breve{\sf K}$
\begin{equation}
  \label{eq:65}
  \breve{\sf K}\equiv\hbar\,{\sf H}^{1/2}\star{\sf \Gamma}\star{\sf
    H}^{1/2}=\breve{\sf K}^\dagger.
\end{equation}
Since $\breve{\sf K}$ is hermitian, one can 
find a star-unitary matrix
${\sf U}$, i.e.
\begin{equation}
  \label{eq:66}
  {\sf U}^\dagger\star{\sf U}={\sf U}\star{\sf U}^\dagger=1,
\end{equation}
which star-diagonalizes $\breve{\sf K}$ according to 
\begin{equation}
  \label{eq:67}
  {\sf U}^\dagger\star\breve{\sf K}\star{\sf U}=\breve{\sf K}_d,
\end{equation}
where $\breve{\sf K}_d$ is a diagonal matrix with real
entries (the `stargenvalues').

Plugging in Eq.~\eqref{eq:65} into Eq.~\eqref{eq:67}, we obtain
\begin{align}
  \label{eq:68}
  \breve{\sf K}_d&=\hbar\,{\sf U}^\dagger\star{\sf H}^{1/2}\star{\sf \Gamma}\star{\sf
                   H}^{1/2}\star{\sf U}\nonumber\\
  &=\hbar\,({\sf U}^\dagger\star{\sf H}^{1/2})\star{\sf \Gamma}\star{\sf H}\star({\sf
                   H}^{-1/2}\star{\sf U}).
\end{align}
Using ${\sf K}=\hbar\,{\sf \Gamma}\star{\sf H}$ we see that this has the form
of Eq.~\eqref{eq:24} if we identify
\begin{equation}
  \label{eq:69}
 \breve{\sf K}_d={\sf K}_d,
\end{equation}
which, in particular, means that ${\sf K}_d$ is real, and
\begin{equation}
  \label{eq:70}
  {\sf S}={\sf H}^{-1/2}\star{\sf U},\quad {\sf S}^{-1}={\sf
    U}^\dagger\star{\sf H}^{1/2},
\end{equation}
up to a gauge choice, as discussed in the main text.

From Eq.~\eqref{eq:70} we can immediately show Eq.~\eqref{eq:26} by
direct calculation.    Eq.~\eqref{eq:26} and Eq.~\eqref{eq:70} can be
considered a choice of normalization of $\mathsf{S}$.  Independent of
that normalization, we note the identity
\begin{equation}
  \label{eq:71}
  {\sf K}^\dagger={\sf H}\star{\sf K}\star{\sf H}^{-1}.
\end{equation}

\subsection{Diagonalization to $O(\hbar)$}
\label{sec:diag-ohbar}

Now we show how to convert the formidable star-diagonalization problem
defined by Eq.~\eqref{eq:24} and Eq.~\eqref{eq:26} into conventional
linear algebra in the semi-classical expansion.  We aim here to obtain
$\mathsf{K}_d$ and $\mathsf{S}$ up to first order in $\shbar$ (and
$\mathsf{S}$ up to gauge freedom, of course).

To do so, we specify the expansions of $\mathsf{K}$ and $\mathsf{S}$.
The former is unambiguously determined by expanding the Moyal product,
regarding $\mathsf{H}$ and
$\mathsf{\Gamma}$ as $\shbar$-independent. We have
\begin{equation}
  \label{eq:72}
  {\sf K}=\hbar\,\left({\sf \Gamma}{\sf
    H}+i\shbar\frac{\epsilon_{\alpha\beta}}{2}\partial_\alpha{\sf
    \Gamma}\partial_\beta{\sf H}\right)+O(\shbar^2),
\end{equation}
and so
\begin{equation}
  \label{eq:73}
  {\sf K}={\sf K}_0+\shbar{\sf K}_1+O(\shbar^2),
\end{equation}
with
\begin{align}
  \label{eq:74}
  {\sf K}_0&=\hbar\,{\sf \Gamma}{\sf H},\nonumber\\
  {\sf K}_1&=i\hbar\frac{\epsilon_{\alpha\beta}}{2}\partial_\alpha{\sf
             \Gamma}\partial_\beta{\sf H}.
\end{align}

Next, we write the expansion of $\mathsf{S}$ according to the
convenient form
\begin{align}
  \label{eq:75}
  \mathsf{S} & = \mathsf{S}_0 \left(1 + \shbar
               \tilde{\mathsf{S}}_1\right) + O(\shbar^2).
\end{align}
Here the first two terms in the expansion are specified by
$\mathsf{S}_0$ and $\tilde{\mathsf{S}}_1$, both $O(\shbar^0)$ a priori, which are to be determined
by enforcing the star-diagonalization conditions up to first order in
$\shbar$.  The hermitian conjugate gives
\begin{align}
  \label{eq:76}
  \mathsf{S}^\dagger & = \left(1 + \shbar
                       \tilde{\mathsf{S}}^\dagger_1\right)\mathsf{S}_0^\dagger
                       + O(\shbar^2).
\end{align}
Inserting Eq.~\eqref{eq:75} and Eq.~\eqref{eq:76} into
Eq.~\eqref{eq:26}, and equating terms at zeroth and first order in
$\shbar$ gives the conditions
\begin{subequations}
\begin{align}
  \mathsf{S}_0^\dagger \mathsf{H} \mathsf{S}_0 & = 1, \label{eq:20a} \\
  \tilde{\mathsf{S}}_1^\dagger + \tilde{\mathsf{S}}_1 &= -
  \frac{i\epsilon_{\alpha\beta}}{2} \left( \partial_\alpha
  \mathsf{S}_0^\dagger \mathsf{H} \partial_\beta \mathsf{S}_0 +
  \mathsf{S}_0^\dagger \partial_\beta\mathsf{H} \mathsf{S}_0+
  \mathsf{S}_0^\dagger \partial_\alpha\mathsf{H} \partial_\beta
  \mathsf{S}_0\right) .  \label{eq:20b}
\end{align}
\end{subequations}
We see that the hermitian part of $\tilde{\mathsf{S}}_1$ is fixed by
the second condition, once the zeroth order term is known.

To proceed, we also need $\mathsf{S}^{-1}$, which we obtain using
Eq.~\eqref{eq:26}, Eq.~\eqref{eq:76} and expanding consistently to $O(\shbar)$:
\begin{align}
  \label{eq:77}
  \mathsf{S}^{-1} & = \mathsf{S}^\dagger \star \mathsf{H} \nonumber \\
  & = \left( 1+ \shbar\left( \tilde{\mathsf{S}}_1^\dagger +
    \frac{i\epsilon_{\alpha\beta}}{2} \partial_\alpha
    \mathsf{S}_0^\dagger \partial_\beta \mathsf{H} \mathsf{S}_0\right)\right)
    \mathsf{S}_0^{-1} + O(\shbar^2)
\end{align}
(take care that $\mathsf{S}^{-1}$ is the
star-inverse not the ordinary matrix inverse).
We also used the zeroth order condition
\begin{align}
  \label{eq:78}
  \mathsf{S}_0^{-1}  & = \mathsf{S}_0^\dagger \mathsf{H},
\end{align}
which follows from Eq.~\eqref{eq:20a}
(here $\mathsf{S}_0^{-1} $ is the $O(\shbar^0)$ term in the expansion of $\ms S^{-1}$
as well as the inverse of $\ms S_0$ for the \emph{usual} matrix product).
Then one can use Eq.~\eqref{eq:20b} to
eliminate $\tilde{\mathsf{S}}_1^\dagger$ in favor of
$\tilde{\mathsf{S}}_1$ within Eq.~\eqref{eq:77}, which yields
\begin{align}
  \label{eq:79}
  \mathsf{S}^{-1} & = \left( 1 - \shbar \left(\tilde{\mathsf{S}}_1 +
                    \frac{i\epsilon_{\alpha\beta}}{2} \partial_\alpha
                    \mathsf{S}_0^{-1} \partial_\beta \mathsf{S}_0
                    \right)\right) \mathsf{S}_0^{-1} +O(\shbar^2) \nonumber \\
  & = \left(1 - \shbar \left( \tilde{\mathsf{S}}_1 -
    \frac{i\epsilon_{\alpha\beta}}{2} \mathsf{\Lambda}_\alpha
    \mathsf{\Lambda}_\beta\right)\right) \mathsf{S}_0^{-1} +O(\shbar^2).
\end{align}

Eq.~\eqref{eq:79} follows from the constraint on ${\sf S}$ from the normalization condition
Eq.~\eqref{eq:26}, but so far we have not used the actual {\em diagonalization} condition in
Eq.~\eqref{eq:24}.  Using Eq.~\eqref{eq:75} and Eq.~\eqref{eq:79} in
Eq.~\eqref{eq:24} and collecting terms to $O(\shbar)$ we find
\begin{align}
  \label{eq:kdd}
  {\sf K}_d= &\, {\sf K}_{0,d}+\shbar \bigg( {\sf S}_0^{-1}{\sf K}_1{\sf
             S}_0+\left[{\sf K}_{0,d},\tilde{\sf
               S}_1\right]\nonumber \\
  &\, -i\frac{\epsilon_{\alpha\beta}}{2}\left\{{\sf
             \Lambda}_\alpha,\partial_\beta{\sf K}_{0,d}-{\sf K}_{0,d}{\sf \Lambda}_\beta\right\}\bigg)+O(\shbar^2),
\end{align}
where
\begin{align}
  \label{eq:80}
  \mathsf{K}_{0,d} & = \mathsf{S}_0^{-1} \mathsf{K}_0 \mathsf{S}_0.
\end{align}
The $O(\shbar^0)$ conditions, 
Eqs.~(\ref{eq:80},\ref{eq:20a}), 
taken together, define a standard (but non-hermitian) eigenvalue problem, where we demand that
$\mathsf{K}_{0,d}$ is diagonal.  By a classical limit, i.e.\ a regular matrix
product version, of the argument
in Sec.~\ref{sec:form-relat-herm}, it is straightforward to show that
a solution always exists and that $\mathsf{K}_{0,d}$ is not only
diagonal but hermitian, i.e.\ real diagonal.

To achieve the star-diagonalization to $O(\shbar)$, we must require
that the term in parenthesis in Eq.~\eqref{eq:kdd} be diagonal,
i.e.\ that
\begin{align}
  \label{eq:81}
  &\bigg( {\sf S}_0^{-1}{\sf K}_1{\sf
             S}_0+\left[{\sf K}_{0,d},\tilde{\sf
  S}_1\right] \\
  &\; -i\frac{\epsilon_{\alpha\beta}}{2}\left\{{\sf
             \Lambda}_\alpha,\partial_\beta{\sf K}_{0,d}-{\sf
  K}_{0,d}{\sf \Lambda}_\beta\right\}\bigg)_{ab}  = 0, && a\neq b.\nonumber
\end{align}
This is a condition on $\tilde{\mathsf{S}}_1$, and while an explicit
form for the latter can be found, we will actually not need it for our calculations, so we suffice to say that Eq.~\eqref{eq:81}
is well-defined and consistent with Eq.~\eqref{eq:20b} and the
hermiticity of $\mathsf{K}_d$.

Since $\mathsf{K}_d$ is diagonal and real, we can take the real part
of the diagonal projection of Eq.~\eqref{eq:kdd}, which gives
\begin{align}
  \label{eq:82}
   {\sf K}_d&={\sf K}_{0,d}+\shbar{\rm Re}\left[\left({\sf S}_0^{-1}{\sf K}_1{\sf
      S}_0\right)^{({\rm
              d})}\right]\\
            & +\shbar\epsilon_{\alpha\beta}\left({\sf A}_\alpha\partial_\beta{\sf K}_{0,d}
              - 
              \frac 1 4 {\rm Im} \left \{ \ms \Lambda_\beta, \left [ \ms \Lambda_\alpha , \ms K_{0,d} \right ] \right \}^{\rm (d)} 
              \right)+O(\shbar^2). \nonumber 
\end{align}

Transforming to the gauge-invariant function
$\underline{\ms K}_d$ then immediately gives
Eq.~\eqref{eq:211a} in the main text, noting that ${\rm Im} \left \{ \ms \Lambda_\beta, \left [ \ms \Lambda_\alpha , \ms K_{0,d} \right ] \right \}^{\rm (d)}$
  is antisymmetric in $\alpha \leftrightarrow \beta$, and so $\sf M_{\alpha\beta} = - \sf M_{\beta\alpha}$.

\subsection{Symmetry of the spectrum}
\label{sec:symmetry-spectrum}

From their definitions, we
know that $\ms \Gamma(X,p)=-\ms \Gamma(X,-p)^*$ and
$\ms H(X,p)= \ms H(X,-p)^*$.  This implies that the dynamical matrix obeys
$\ms K(X,p)=-\ms K(X,-p)^*$.  The latter induces relations amongst the
star-eigenvalues of $\mathsf{K}_d$ at reversed momenta.  Taking the
complex conjugate of Eq.~\eqref{eq:24} evaluated at momentum $-p$, and
using the fact that $\mathsf{K}_d$ is real, we deduce that
\begin{align}
  \label{eq:132}
  \mathsf{K}_d(X,-p) =- \left[\mathsf{S}^{-1}(X,-p)\right]^*\star \mathsf{K}(X,p)
  \star \left[\mathsf{S}(X,-p)\right]^*.
\end{align}
Taking the complex conjugate of Eq.~\eqref{eq:26} at momentum $-p$ gives
\begin{align}
  \label{eq:134}
  \mathsf{S}(X,-p)^\top \star \mathsf{H}(X,p) \star \mathsf{S}(X,-p)^* = 1.
\end{align}
These two equations are solved by using again Eqs.~(\ref{eq:24},\ref{eq:26}), by
\begin{align}
  \label{eq:135}
  \mathsf{S}(X,-p) = \mathsf{S}(X,p)^*, && \mathsf{K}_d(X,-p) = - \mathsf{K}_d(X,p),
\end{align}
\emph{up to a gauge transformation}, and up to permutations of the
diagonal entries of $\mathsf{K}_d$.  We see that the negative energies
at momentum $p$ appear with the opposite sign as positive energies at
momentum $-p$, and vice versa.  Thus the modes of the spectrum of $\ms K$ come in positive (the
``physical'' ones) and negative energy pairs, $(\varepsilon_a(X,p),
\varepsilon_{\bar{a}}(X,p)\equiv-\varepsilon_a(X,-p))$. This defines
the notation $\bar{a}$.   Moreover, we can deduce that
\begin{align}
  \label{eq:133}
  [\ms \Omega_{\alpha\beta}(X,-p)]_{\bar a \bar a} &=  - [\ms \Omega_{\alpha\beta}(X,p)]_{a a} ,\nonumber \\
   [\ms M_{\alpha\beta}(X,-p)]_{\bar a \bar a} &= + [\ms M_{\alpha\beta}(X,p)]_{a a} .
\end{align}
We make use of these last two properties in Sec.~\ref{sec:unif-finite-syst} and App.~\ref{sec:discussion--currents}.

\section{Diagonalization of $\mathcal{J}_\alpha$}
\label{sec:diagonalization}

Here we provide details of the calculations leading from
Eq.~\eqref{eq:23} to Eqs.~(\ref{eq:565},\ref{eq:66a},\ref{eq:66b}) in
Sec.~\ref{sec:energy-current}.

Beginning with Eq.~\eqref{eq:23}, we use
Eq.~\eqref{eq:51} and Eq.~\eqref{eq:53} to obtain
\begin{align}
  \label{eq:83}
  \mathcal{J}_\alpha
  &=\frac{\upsilon}{2}\epsilon_{\alpha\beta}{\rm Re}{\rm
    Tr}\Big[\left({\sf S}\star{\sf Q}_\beta\star{\sf S}^{-1}\right)\left({\sf S}\star{\sf
    F}_d\star{\sf S}^{-1}\right)\Big]+O(\upsilon\shbar^2).
\end{align}
We now use Eq.~\eqref{eq:62} with ${\sf O}_1={\sf S}\star{\sf
  Q}_\beta\star{\sf S}^{-1}$ and ${\sf O}_2={\sf S}\star{\sf
  F}_d\star{\sf S}^{-1}$ to ``restore'' a star product and as a second step
keep only the zeroth order term of the expansion of the star products
in terms which are already of order $\shbar$ (which amounts to
turning the star products into regular matrix products):
\begin{widetext}
  \begin{align}
    \label{eq:84}
  \mathcal{J}_\alpha&=\frac{\upsilon}{2}\epsilon_{\alpha\beta}{\rm Re}{\rm
    Tr}\Big[{\sf S}\star{\sf Q}_\beta\star{\sf
    F}_d\star{\sf S}^{-1}\Big]
    +\frac{\upsilon\shbar}{4}\epsilon_{\alpha\beta}\epsilon_{\gamma\lambda}{\rm Im}{\rm
    Tr} \left[\partial_\gamma \left({\sf S}{\sf Q}_\beta{\sf S}^{-1}\right)\partial_\lambda\left({\sf S}{\sf
    F}_d{\sf S}^{-1}\right)\right]+O(\upsilon\shbar^2).
  \end{align}
Now we use Eq.~\eqref{eq:46} with ${\sf O}={\sf Q}_\beta\star{\sf
  F}_d$, extract a $\partial_\gamma$ derivative using
$\epsilon_{\gamma\lambda}\partial^2_{\gamma\lambda}=0$, expand the
$\partial_\lambda$ derivative using the chain rule, and use the
cyclicity of the trace:
\begin{align}
  \label{eq:85}
 \mathcal{J}_\alpha &=\frac{\upsilon}{2}\epsilon_{\alpha\beta}{\rm Re}{\rm
    Tr}\Big[{\sf Q}_\beta\star{\sf F}_d\Big]-\frac{\upsilon}{2}\shbar\epsilon_{\alpha\beta}\epsilon_{\gamma\lambda}\partial_\lambda{\rm
    Im}{\rm Tr}[{\sf S}^{-1}\partial_\gamma{\sf S}\left({\sf
    Q}_\beta{\sf F}_d\right)]\nonumber\\ &\qquad+\frac{\upsilon\shbar}{4}\epsilon_{\alpha\beta}\epsilon_{\gamma\lambda}\partial_\gamma{\rm Im}{\rm
    Tr} \left[{\sf S}{\sf Q}_\beta{\sf S}^{-1}\left(\partial_\lambda{\sf S}{\sf
    F}_d{\sf S}^{-1}+{\sf S}\partial_\lambda{\sf F}_d{\sf S}^{-1}+{\sf
    S}{\sf F}_d\partial_\lambda{\sf S}^{-1}\right)\right]+O(\upsilon\shbar^2)\nonumber\\
 &=\frac{\upsilon}{2}\epsilon_{\alpha\beta}{\rm Re}{\rm
    Tr}\Big[{\sf Q}_\beta\star{\sf F}_d\Big]+\frac{\upsilon\shbar}{4}\epsilon_{\alpha\beta}\epsilon_{\gamma\lambda}\partial_\gamma{\rm Im}{\rm
    Tr} \left[ {\sf Q}_\beta\left(\left\{{\sf S}^{-1}\partial_\lambda{\sf S},{\sf
    F}_d\right\}+\partial_\lambda{\sf F}_d\right)\right]+O(\upsilon\shbar^2).
\end{align}
Then, plugging in the expression for ${\sf Q}_\beta$ in Eq.~\eqref{eq:52}, we get
\begin{align}
  \label{eq:86}
  \mathcal{J}_\alpha&=\frac{\upsilon}{2}\epsilon_{\alpha\beta}\left({\rm Re}{\rm
    Tr}\Big[\partial_\beta{\sf K}_d\star{\sf F}_d\Big]+{\rm Re}{\rm
    Tr}\Big[[{\sf S}^{-1}\star\partial_\beta{\sf
                      S}\,\overset{\star},\,{\sf K}_d]\star{\sf
                      F}_d\Big]\right)\nonumber\\
  &\quad+\frac{\upsilon\shbar}{4}\epsilon_{\alpha\beta}\epsilon_{\gamma\lambda}\partial_\gamma{\rm Im}{\rm
    Tr} \left[ (\partial_\beta{\sf K}_d+[{\sf
    S}^{-1}\partial_\beta{\sf S},{\sf K}_d])\left(\partial_\lambda{\sf F}_d+\left\{{\sf S}^{-1}\partial_\lambda{\sf S},{\sf
    F}_d\right\}\right)\right]+O(\upsilon\shbar^2).
\end{align}
In turn, expanding the star products yields
\begin{align}
  \label{eq:87}
  \mathcal{J}_\alpha&=\frac{\upsilon}{2}\epsilon_{\alpha\beta}\left({\rm Re}{\rm
    Tr}\Big[\partial_\beta{\sf K}_d{\sf F}_d\Big]+{\rm Re}{\rm
    Tr}\Big[[{\sf S}^{-1}\partial_\beta{\sf
                      S},{\sf K}_d]{\sf
                      F}_d\Big]\right)\nonumber\\
  &\quad-\frac{\upsilon\shbar}{4}\epsilon_{\alpha\beta}\epsilon_{\gamma\lambda}{\rm Im}{\rm
    Tr} \left[\partial^2_{\beta\gamma}{\sf K}_d\partial_\lambda{\sf
    F}_d+\partial_\gamma[{\sf S}^{-1}\partial_\beta{\sf S},{\sf K}_d]\partial_\lambda{\sf F}_d+\{\partial_\gamma({\sf
    S}^{-1}\partial_\beta{\sf S}),\partial_\lambda{\sf K}_d\}{\sf
    F}_d+[\partial_\gamma{\sf S}^{-1}\partial^2_{\beta\lambda}{\sf
    S},{\sf K}_d]{\sf F}_d\right]\nonumber\\
&\quad+\frac{\upsilon\shbar}{4}\epsilon_{\alpha\beta}\epsilon_{\gamma\lambda}\partial_\gamma{\rm Im}{\rm
    Tr} \left[ (\partial_\beta{\sf K}_d+[{\sf
    S}^{-1}\partial_\beta{\sf S},{\sf K}_d])\left(\partial_\lambda{\sf F}_d+\left\{{\sf S}^{-1}\partial_\lambda{\sf S},{\sf
    F}_d\right\}\right)\right]+O(\upsilon\shbar^2).
\end{align}
The contributions from $\partial^2_{\beta\gamma}{\sf K}_d\partial_\lambda{\sf
    F}_d$ and $\partial_\beta{\sf K}_d\partial_\lambda{\sf F}_d$
  vanish because the latter are real, and those involving commutators,
  except for $[{\sf
    S}^{-1}\partial_\beta{\sf S},{\sf K}_d]\left\{{\sf S}^{-1}\partial_\lambda{\sf S},{\sf
    F}_d\right\}$, also vanish because the trace selects only the
diagonal elements. One is then left with
\begin{align}
  \label{eq:88}
  \mathcal{J}_\alpha&=\frac{\upsilon}{2}\epsilon_{\alpha\beta}{\rm Re}{\rm
                      Tr}\Big[\partial_\beta{\sf K}_d{\sf F}_d\Big]
                      -\frac{\upsilon\shbar}{4}\epsilon_{\alpha\beta}\epsilon_{\gamma\lambda}{\rm Im}{\rm
    Tr} \Big[\{\partial_\gamma \mathsf{\Lambda}_\beta
                                    ,\partial_\lambda{\sf K}_d\}{\sf
                                    F}_d\Big]\nonumber\\
 &\quad+\frac{\upsilon\shbar}{4}\epsilon_{\alpha\beta}\epsilon_{\gamma\lambda}\partial_\gamma{\rm Im}{\rm
    Tr} \Big[ [\mathsf{\Lambda}_\beta,{\sf K}_d]\left\{\mathsf{\Lambda}_\lambda,{\sf
    F}_d\right\}+\partial_\beta{\sf K}_d\left\{\mathsf{\Lambda}_\lambda,{\sf
    F}_d\right\}\Big]+O(\upsilon\shbar^2).
\end{align}
Here we replaced $\mathsf{S}^{-1} \partial_\rho \mathsf{S}
\rightarrow \mathsf{\Lambda}_\rho$ which is correct to this order.

In almost all terms in Eq.~\eqref{eq:88}, the imaginary diagonal part
of $\mathsf{\Lambda}_\rho$, which is simply $\mathsf{A}_\rho$, is selected by the trace against other
diagonal real functions.  The one term not of this form can be
rewritten 
using the cyclicity of the trace as
\begin{align}
  \label{eq:89}
  {\rm Tr}\left([\mathsf{\Lambda}_\alpha ,{\sf K}_d]\{\mathsf{\Lambda}_\beta,{\sf F}_d\}\right)&={\rm
    Tr}\left(\left\{\mathsf{\Lambda}_\beta,\left[\mathsf{\Lambda}_\alpha,{\sf
                                                                                                 K}_d\right]\right\}{\sf
                                                       F}_d\right)=2{\rm Tr}\left(\mathsf{M}_{\alpha\beta}{\sf F}_d\right),
\end{align}
where $\mathsf{M}_{\alpha\beta}$ defined in Eq.~\eqref{eq:61a}.  
Consequently,
\begin{align}
  \label{eq:90}
  \mathcal{J}_\alpha&=\frac{\upsilon}{2}\epsilon_{\alpha\beta}\left({\rm
    Tr}\Big[\partial_\beta{\sf K}_d{\sf F}_d\Big]+\frac{\shbar}{2}\epsilon_{\gamma\lambda}\left({\rm
    Tr} \left[-\{\partial_\gamma{\sf A}_\beta,\partial_\lambda{\sf K}_d\}{\sf
    F}_d+ \partial_\gamma(\partial_\beta{\sf K}_d\left\{{\sf A}_\lambda,{\sf
    F}_d\right\})\right]+2\partial_\gamma{\rm
    Tr} \left[ \mathsf{M}_{\beta\lambda}{\sf
    F}_d\right]\right)\right)+O(\upsilon\shbar^2)\nonumber\\
  &=\frac{\upsilon}{2}\epsilon_{\alpha\beta}\left({\rm
    Tr}\Big[\partial_\beta{\sf K}_d{\sf F}_d\Big]+\shbar\epsilon_{\gamma\lambda}\left({\rm
    Tr} \left[\partial_\gamma({\sf A}_\lambda\partial_\beta{\sf
    K}_d-{\sf A}_\beta\partial_\lambda{\sf K}_d){\sf
    F}_d+ {\sf A}_\lambda \partial_\beta{\sf K}_d\partial_\gamma{\sf
    F}_d\right]+\partial_\gamma{\rm
    Tr} \left[ \mathsf{M}_{\beta\lambda}{\sf F}_d\right]\right)\right)+O(\upsilon\shbar^2).
\end{align}
From here, it is straightforward to separate the term containing
$\mathsf{M}_{\beta\lambda}$ from the others, and with
some rearrangement to first order in $\shbar$, obtain
Eqs.~(\ref{eq:565},\ref{eq:66a},\ref{eq:66b}) in Sec.~\ref{sec:energy-current}.

\end{widetext}

\section{Jacobian}
\label{sec:jacobean}

When expressed in gauge invariant band diagonal form, physical
quantities like the energy and current density display explicit
dependence on the ``Jacobian'' factor defined in Eq.~\eqref{eq:14}.
Here we show that this can indeed be regarded as a Jacobian, i.e.\ a
determinant defining the measure for a change of variables in phase
space.

Consider the change of variables in Eq.~\eqref{eq:10} from
$\ul{\smash{q}}$ to $q$, which induces the change in measure
\begin{align}
  \label{eq:91}
  d^{D}\ul{\smash{q}} & = d^Dq\, \left|\textrm{det}_{D} \left[\left(\frac{\partial\ul{\smash{q}}_\alpha}{\partial
      q_\beta}\right)_{\alpha\beta}\right]\right|.
\end{align}
Here the measure is defined in phase space, with total dimension $D=2d$
twice the spatial dimension, and the subscript $D$ on the determinant
is meant to indicate that the determinant is in this space.  One
should avoid confusion with band space, which plays a trivial role
here because the change of variables is diagonal (since $\mathsf{A}$
in Eq.~\eqref{eq:10} is diagonal).  The determinant can be regarded
as a scalar for a given band, or a diagonal matrix in band space, evaluated
separately for each band.  

The Jacobian of the transformation is just this determinant.  Using
the explicit form of Eq.~\eqref{eq:10},
\begin{align}
  \label{eq:92}
  \mathfrak{J} & = {\rm
    det}_D\left[\left(\frac{\partial\ul{\smash{q}}_\alpha}{\partial
                 q_\beta}\right)_{\alpha\beta}\right] \\
  &={\rm
det}_D\left[\left(\delta_{\alpha\beta}+\shbar\epsilon_{\alpha\gamma}\partial_\beta{\sf
    A}_\gamma\right)_{\alpha\beta}\right]+O(\shbar^2). \nonumber
\end{align}
Note that we have dropped the absolute value of the determinant,
because the Jacobian is perturbatively close to $1$ in the
semi-classical expansion and so has definite sign.  
We have, with ${\rm Tr}_D$ the trace in phase space,
 \begin{align}
   \label{eq:93}
    \mathfrak{J} 
    & =e^{{\rm Tr}_D[\ln(\left(\delta_{\alpha\beta}+\shbar\epsilon_{\alpha\gamma}\partial_\beta{\sf
                                           A}_\gamma\right)_{\alpha\beta})]}+O(\shbar^2)\nonumber\\
   & =e^{{\rm Tr}_D[\left(\shbar\epsilon_{\alpha\gamma}\partial_\beta{\sf
     A}_\gamma\right)_{\alpha\beta}]}+O(\shbar^2)\nonumber\\
  & =1+{\rm Tr}_D[\left(\shbar\epsilon_{\alpha\gamma}\partial_\beta{\sf
     A}_\gamma\right)_{\alpha\beta}]+O(\shbar^2)\nonumber\\
  & =1+\shbar\epsilon_{\alpha\beta}\partial_\alpha{\sf
     A}_\beta +O(\shbar^2).
 \end{align}
 This is precisely the definition in Eq.~\eqref{eq:14}.

\section{Explicit separation into position and momentum coordinates}
\label{sec:expl-separ-into}

In the main text, we presented the quasiparticle equations of motion
and current density in compact forms combining positions and momenta
into a single phase space coordinate.  To connect to more familiar
forms, we expand these explicitly here.

\begin{widetext}
Writing $q_{X_\mu} = X_\mu$ and $q_{p_\mu} = p_\mu$, we obtain from Eq.~\eqref{eq:196} the two
analogs of Hamilton's equations 
\begin{subequations}
  \begin{align}
  \label{eq:192a}
   \partial_t{\sf
  X}_\mu&=\partial_{p_\mu}
          \ul{\sf K}_d-\shbar\left({\sf
    \Omega}_{p_\mu p_\nu}\partial_{X_\nu}\ul{\sf K}_d -{\sf
                         \Omega}_{p_\mu X_\nu}\partial_{p_\nu}\ul{\sf K}_d\right)+O(\shbar^2),\\
    \label{eq:192b}
    \partial_t{\sf p}_\mu&=- \partial_{X_\mu}\ul{\sf K}_d+\shbar\left({\sf
    \Omega}_{X_\mu p_\nu}\partial_{X_\nu}\ul{\sf K}_d -{\sf
    \Omega}_{X_\mu X_\nu}\partial_{p_\nu}\ul{\sf K}_d\right)+O(\shbar^2).
\end{align}
\end{subequations}
This defines a velocity $\mathsf{v}_\mu \equiv \partial_t {\sf
  X}_\mu$ and a force $\mathsf{f}_\mu \equiv \partial_t{\sf
  p}_\mu$.

We can similarly inspect the real space and momentum space components
of $\mathcal{J}_\alpha$. Taking $\alpha = X_\mu$, we obtain the
quantity describing the flow of energy in the $X_\mu$ direction.  This
is just the momentum-resolved energy current.  We have
\begin{align}
  \label{eq:94}
  \mathcal{J}_{X_\mu}&=\frac{\upsilon}{2}{\rm
    Tr}\left[\,\mathfrak{J}\,\partial_t {\sf X}_\mu\, \ul{\sf F}_d+\shbar\left(
      \partial_{X_\nu}\left(\mathsf{M}_{p_\mu p_\nu}\ul{\sf F}_d\right)-\partial_{p_\nu}\left(\mathsf{M}_{p_\mu X_\nu}\ul{\sf F}_d\right)\right)\right]+O(\upsilon\shbar^2)\nonumber\\
  &=\frac{\upsilon}{2}{\rm
    Tr}\left[\,\mathfrak{J}\,\partial_t {\sf X}_\mu\, \ul{\sf F}_d+\shbar\epsilon^{\mu\nu\lambda}\left(\partial_{X_\nu}\mathfrak{M}_{pp}^\lambda
    -\partial_{p_\nu}\mathfrak{M}_{pX}^\lambda\right)\right]+O(\upsilon\shbar^2),
\end{align}
where $\epsilon^{\mu\nu\lambda}$ (note here we use superscripts
and not subscripts for the indices) is the usual 3d Levi-Civita tensor, and
\begin{equation}
  \label{eq:95} \mathfrak{M}_{pp}^\lambda=\frac{1}{2}\epsilon^{\mu\nu\lambda}\mathsf{M}_{p_\mu p_\nu}\ul{\sf F}_d,
  \qquad \mathfrak{M}_{pX}^\lambda=\frac{1}{2}\epsilon^{\mu\nu\lambda}\mathsf{M}_{p_\mu X_\nu}\ul{\sf F}_d,
\end{equation}
because $\mathsf{M}_{\alpha\beta}=-\mathsf{M}_{\beta\alpha}$.
The total energy current at position $X_\mu$ is obtained by
integrating $\mathcal{J}_{X_\mu}(X_\mu,p_\mu)$ over $p$.
Then clearly the $\partial_{p_\nu}\mathfrak{M}_{pX}^\lambda$ term drops.
  Moreover, if one computes the flux of $\mathcal{J}_{X_\mu}$ through an open 2d surface
  at whose boundary $\mathfrak{M}_{pp}^\lambda$ vanishes, the $\partial_{X_\nu}\mathfrak{M}_{pp}^\lambda$ term drops.

Taking now $\alpha = p_\mu$ describes the flow of energy in momentum
space.  This is something like an ``energy force'' density.  It is
\begin{align}
  \label{eq:96}
    \mathcal{J}_{p_\mu}&=\frac{\upsilon}{2}{\rm
    Tr}\left[\,\mathfrak{J}\,\partial_t {\sf p}_\mu\, \ul{\sf F}_d-\shbar\left(
      \partial_{X_\nu}\left(\mathsf{M}_{X_\mu p_\nu}\ul{\sf F}_d\right)-\partial_{p_\nu}\left(\mathsf{M}_{X_\mu X_\nu}\ul{\sf F}_d\right)\right)\right]+O(\upsilon\shbar^2)\nonumber\\
  &=\frac{\upsilon}{2}{\rm
    Tr}\left[\,\mathfrak{J}\,\partial_t {\sf p}_\mu\, \ul{\sf F}_d-\shbar\epsilon^{\mu\nu\lambda}\left(\partial_{X_\nu}\mathfrak{M}_{Xp}^\lambda-\partial_{p_\nu}\mathfrak{M}_{XX}^\lambda\right)\right]+O(\upsilon\shbar^2),
\end{align}
where
\begin{equation}
  \label{eq:97}
  \mathfrak{M}_{Xp}^\lambda=\frac{1}{2}\epsilon^{\mu\nu\lambda}\mathsf{M}_{X_\mu p_\nu}\ul{\sf F}_d,\qquad \mathfrak{M}_{XX}^\lambda=\frac{1}{2}\epsilon^{\mu\nu\lambda}\mathsf{M}_{X_\mu X_\nu}\ul{\sf F}_d.
\end{equation}
\end{widetext}

\section{The case of a separable position and momentum dependence}
\label{sec:case-grav-field}

Here we specialize our theory to the case where the dependence on $X$
and $p$ of ${\sf H}$ is separable, i.e.\ we can write ${\sf
  H}(X,p)=g(X){\sf H}^{\rm h}(p)$, where $g$ is a c-number real function, and
the superscript ${\rm h}$ stands for ``homogeneous,'' and we
additionally require ${\sf \Gamma}(X,p)={\sf \Gamma}(p)={\sf
  \Gamma}^{\rm h}(p)$. In other words, all the spatial dependence is
encoded in the function $g(X)$.

\subsection{Solving the inhomogeneous problem}
\label{sec:solv-inhom-probl}

In this section, we show how the solution to the inhomogeneous problem Eqs.~(\ref{eq:24},\ref{eq:26})
can be deduced from that of the homogeneous one, Eqs.~(\ref{eq:24hom},\ref{eq:26hom}) in the main text.

This can be worked out perturbatively in $\shbar$.
To zeroth order, comparing Eq.~\eqref{eq:26hom} to Eq.~\eqref{eq:26} yields immediately
\begin{align}
  \label{eq:104}
  \ms S_0 &= \ms S^{\rm h}/\sqrt g ,
\end{align}
whence
\begin{align}
  \label{eq:103}
  \ms \Lambda_\alpha &= - \frac {\partial_\alpha g}{2 g} + (\ms S^{\rm h})^{-1}\partial_\alpha \ms S^{\rm h} + O(\shbar) .
\end{align}
Explicitly in terms of position and momentum components, and up to corrections $O(\shbar)$,
this reads
$ \ms \Lambda_{p_\mu}= \ms \Lambda^{\rm h}_{p_\mu}$ so that $ \ms A_{p_\mu}= \ms A^{\rm h}_{p_\mu}$,
and $ \ms \Lambda_{X_\mu}= - \frac 1 2 \psi_\mu$ and in turn $\ms A_{X_\mu}= 0$.
Here we defined $\psi_\mu = \partial_{X_\mu}g /g$ which is sometimes called
a ``gravitational field'' 
\cite{luttinger1964theory,cooper1997,qin2012,matsumoto2014}.

The decomposition Eq.~\eqref{eq:74}, ${\sf K}={\sf K}_0+\shbar{\sf K}_1+O(\shbar^2)$, becomes here
\begin{align}
  \label{eq:105}
  \ms K_{0} &= g \ms K^{\rm h}, \nonumber\\
  \ms K_1 &= \hbar\,\frac i 2 \partial_{X_\mu}g \partial_{p_\mu} \ms \Gamma^{\rm h} \ms H^{\rm h}.
\end{align}
Thus $\ms K_{0,d}= g \ms K_{d}^{\rm h}$.
Then one just need to plug Eqs.~(\ref{eq:104}, \ref{eq:105}) into Eq.~\eqref{eq:82}.
Because ${\sf S}_0^{-1}{\sf K}_1{\sf S}_0$ is purely anti-hermitian,
its real diagonal part vanishes, so the first $O(\shbar)$ term in
Eq.~\eqref{eq:82}, ${\rm Re}\left[\left({\sf S}_0^{-1}{\sf K}_1{\sf S}_0\right)^{({\rm d})}\right]$, is zero.
Besides, because of the $\alpha \leftrightarrow \beta$ antisymmetry, for the last term it is sufficient to look at
$\alpha=X_\mu,\beta = p_\mu$, where clearly the commutator vanishes because $\ms \Lambda_{X_\mu}$ is diagonal (actually $\propto 1$),
so the third $O(\shbar)$ term, $\epsilon_{\alpha\beta}{\rm Im}\{{\sf
  \Lambda}_\beta,[{\sf \Lambda}_\alpha,{\sf K}_{0,d}]\}^{({\rm d})}$, is zero as well.
Consequently, the only $O(\shbar^1)$ contribution in Eq.~\eqref{eq:82} comes from the second $O(\shbar^1)$ term, and
\begin{align}
  \label{eq:107}
\ms K_d &= g \left ( 1 + \shbar \psi_\mu \ms A_{p_\mu}^{\rm h} \right ) \ms K_{d}^{\rm h} +O(\shbar^2).
\end{align}
This solves the problem to order $O(\shbar^1)$.

\subsection{Distribution function}
\label{sec:distr-funct}

The distribution $\ms F_d$ close to equilibrium is a function of the
energies ${\sf K}_d$ only,
\begin{align}
  \label{eq:108}
  \ms F_d &= f({\sf K}_{d},T) \\
  & = f( g{\sf K}_{d}^{\rm h},T) + g \psi_\mu \ms A_{p_\mu}^{\rm h}
    {\sf K}_{d}^{\rm h}  f'(g{\sf K}_{d}^{\rm h},T) +O(\shbar^2), \nonumber 
\end{align}
where a function of a diagonal matrix is transparently defined (cf.\ Eq.~\eqref{eq:112a}).
It should be understood that $f$ is a real function which depends solely on \emph{local equilibrium} properties,
such as the local temperature \cite{cooper1997,qin2011energy}, and
$f'$ is the derivative of $f$ with respect to its first variable. 
Because $f$ does not depend on the profile of $g(X)$, it is the same function as in the homogeneous case,
$\ms F_d^{\rm h}=f({\sf K}_d^{\rm h},T)$. It is then not difficult to show that
\begin{align}
  \label{eq:109}
  f(\varepsilon,T) &= \varepsilon \left ( n_{\rm B}(\varepsilon,T)+ \tfrac 1 2 \right ),
\end{align}
where $n_{\rm B}(\varepsilon,T) = [ \exp(\varepsilon/k_{\rm B}T)-1 ]^{-1}$ is the Bose function, $T$ is the local temperature,
and the extra $\varepsilon$ factor comes from the normalization of eigenvectors, Eq.~\eqref{eq:26hom}.

\subsection{Energy current}
\label{sec:energy-current-1}

We are now in a position to compute the energy current,
Eq.~\eqref{eq:565}. First consider Eq.~\eqref{eq:66a}. Because the
only non-vanishing curvature here is $\ms \Omega_{p_\mu p_\nu}$ since
${\sf A}_\mu=0$,
the jacobian factor reduces to $\mf J=1+O(\shbar^2)$.
We specialize to $\beta=p_\mu$, and only $\sigma = X_\nu$ in the second factor can contribute (cf.\ Eq.~\eqref{eq:192a}), yielding
\begin{align}
  \label{eq:110}
  \partial_t \ms X_\mu &=  \partial_{p_\mu} \ul{\sf K}_d -\shbar {\sf
                         \Omega}_{p_\mu p_\nu}\partial_{X_\nu}\ul{\sf K}_d +O(\shbar^2) \nonumber \\
  &= g \left ( \partial_{p_\mu} {\sf K}_d^{\rm h} - \shbar {\sf
    \Omega}^{\rm h}_{p_\mu p_\nu} \psi_\nu {\sf K}_d^{\rm h} \right )  +O(\shbar^2) ,
\end{align}
where the $O(\shbar)$ terms generated by going from $\ul{\sf K}_d$ to
${\sf K}_d$
canceled against each other.
Finally, plugging Eq.~\eqref{eq:108} into the third factor and
expanding, it is straightforward to show that the latter reduces to
$f(g {\sf K}_d^{\rm h},T)$.
The first contribution to the current is thus
\begin{align}
  \label{eq:111}
  \mathcal{J}_{X_\mu} ^{(1)}&=\frac{\upsilon}{2}{\rm Tr}\Big[ g \left (
                              \partial_{p_\mu} {\sf K}_d^{\rm h}
                              - \shbar {\sf \Omega}^{\rm h}_{p_\mu
                              p_\nu} \psi_\nu {\sf K}_d^{\rm h} \right
                              ) f(g {\sf K}_d^{\rm h},T) \Big]+O(\upsilon\shbar^2). 
\end{align}
Note that the first (order $O(\shbar^0)$) contribution is a total momentum derivative,
therefore it always vanishes after momentum integration.

Now consider Eq.~\eqref{eq:66b}. Because $\sf M_{\alpha\beta}$ is antisymmetric and $\ms \Lambda_{X_\mu}$ is diagonal,
the only (to $O(\shbar^0)$) nonvanishing component of $\sf M$ is 
\begin{align}
  \label{eq:112}
  {\sf M}_{p_\mu p_\nu} &= g \frac{1}{2}{\rm Im}
  \left\{ \mathsf{\Lambda}_{p_\nu}^{\rm h}
                          ,\left[\mathsf{\Lambda}_{p_\mu}^{\rm h}
                          ,{\sf K}_d^{\rm h} \right] \right\}^{({\rm d})} + O(\shbar),
\end{align}
whence the second contribution to the current
\begin{align}
  \label{eq:113}
  \mathcal{J}_{X_\mu}^{(2)}&=\frac{\upsilon \shbar}{2} \partial_{X_\nu}{\rm Tr}\big[\mathsf{M}_{p_\mu p_\nu} {\sf F}_d \big] + O(\upsilon\shbar^2) \\
                           &= \frac {\upsilon \shbar} 2 {\rm   Tr}\bigg [ g\mathsf{M}_{p_\mu p_\nu}^{\rm h}
                             \Big ( f(g {\sf K}_d^{\rm h},T) \left (  \psi_\nu + \partial_{\nu}T / T \right )  \nonumber \\
  & + g  {\sf K}_d^{\rm h} f'(g {\sf K}_d^{\rm h},T) \left (  \psi_\nu -  \partial_{\nu}T / T \right ) \Big ) \bigg ] + O(\upsilon\shbar^2) .\nonumber
\end{align}

Eqs.~(\ref{eq:111},\ref{eq:113}) provide the total local,
momentum-resolved energy current.

\subsection{Discussion}
\label{sec:discussion--currents}

Here we restrict ourselves to the equilibrium case, $\partial_{X_\nu}T = 0$.
Then $ \mathcal{J}_{X_\mu} ^{(1)}$ besides  $\mathcal{J}_{X_\mu} ^{(2)}$ is also a total $\partial_{X_\nu}$ derivative,
so that the net energy current $\bs J_{\rm tot}$ vanishes, as it should in equilibrium.
Meanwhile, neither the \emph{local} momentum-integrated currents 
\begin{align}
  {J}_{X_\mu} ^{(1)} &= - \upsilon \shbar  g \;\psi_\nu \int_p {\rm Tr}_+ \Big[ {\sf \Omega}^{\rm h}_{p_\mu p_\nu} {\sf K}_d^{\rm h} \, f(g {\sf K}_d^{\rm h},T) \Big] + O(\upsilon\shbar^2), \nonumber\\
  {J}_{X_\mu}^{(2)} &= \upsilon \shbar \;\partial_{X_\nu} \int_p {\rm Tr}_+  \Big[ g\mathsf{M}^{\rm h}_{p_\mu p_\nu} \,f(g {\sf K}_d^{\rm h},T)  \Big ] + O(\upsilon\shbar^2)  ,   \label{eq:130}
\end{align}
nor their sum $J_{X_\mu}(X) \equiv  {J}_{X_\mu} ^{(1)} +{J}_{X_\mu} ^{(2)}$ need vanish even at equilibrium,
because the energy magnetization is not zero a priori. Note that in Eqs.\eqref{eq:130} the trace $\rm Tr_+$ runs over positive energy eigenstates only,
following the argument developed in Sec.~\ref{sec:unif-finite-syst} and App.~\ref{sec:symmetry-spectrum}.

There is however a physical constraint that the energy magnetization should obey the third law of thermodynamics,
so that $J_{X_\mu}(X) $ should at least vanish in the limit of zero temperature. Let us show that Eqs.~(\ref{eq:111},\ref{eq:113}) indeed satisfy this property.
We will resort to the fact that $f(\epsilon,T)\overset{T \rightarrow 0} = \frac 1 2 \epsilon$ for any $\epsilon > 0$,
in other words only the zero-point fluctuations contribute to the density matrix.
Taking this limit within Eq.\eqref{eq:130}, expanding the definition Eq.\eqref{eq:61a},
using cyclicity of the trace and eventually identifying $\ms \Omega_{p_\mu p_\nu}^{\rm h}={\rm Im}[\ms \Lambda_{p_\nu}^{\rm h}, \ms \Lambda_{p_\mu}^{\rm h}]$,
it is then straightforward to show that (in equilibrium)
\begin{align}
  {J}_{X_\mu}^{(2)} \quad& \overset{T \rightarrow 0} =\quad  \frac{\upsilon\shbar} 2   g^2 \psi_\nu \int_p
                      {\rm Tr}_+  \left ( \ms K_d^{\rm h} \ms \Omega_{p_\mu p_\nu}^{\rm h} \ms K_d^{\rm h} \right ) + O(\upsilon\shbar^2) \nonumber \\
                      &\overset{T \rightarrow 0} = \quad -  {J}_{X_\mu}^{(1)} \;+ O(\upsilon\shbar^2) \label{eq:129}.
\end{align}
This shows, to $O(\upsilon\shbar^1)$, the cancellation of local magnetization
currents in the zero-temperature limit (this is in fact already true for the momentum-resolved current).
More generally, the above shows that the zero-point fluctuations (the `$+ \frac 1 2$' in the definition of $f$)
  cancel exactly between $ {\mc J}_{X_\mu}^{(1)} $ and $ {\mc J}_{X_\mu}^{(2)} $.

\begin{widetext}
  
\section{Explicit spatial integration for a homogeneous system with a boundary}
\label{sec:expl-spat-integr}

We now perform the $\int \text dx$ (recall $x\equiv X_x$) integration in Eqs.~(\ref{eq:115},\ref{eq:116}).
We have
\begin{align}
  \label{eq:118}
  \kappa_{xy}^{\rm tr} &= -\frac {1} {\partial_x T} {\rm Tr}_+ \int_p \lim_{L\rightarrow \infty} \frac 1 L\int \text dx \,
                         {\sf \Omega}^{\rm h}_{p_y p_x} \partial_x
                         \left ( {\sf K}_d(x) \right ) {\sf K}_d(x)
                         \left ( n_{\rm B}({\sf K}_d(x),T) \big |_{T(x)} + \tfrac 1 2 \right ) ,
\end{align}
where we recall ${\sf K}_d(x) =g(x) {\sf K}_d^{\rm h}$.
Clearly the integrand decays over distances $O(L^0)$ outside of $I_x$,
therefore one does not change the result by multiplying the former by any cutoff function $\Lambda(x)$ strictly equal to 1 wherever the integrand is nonzero,
and decaying slowly to zero at $x\rightarrow\pm\infty$.
In addition we introduce $1 = \int \text d \varepsilon
\,\delta(\varepsilon-{\sf K}_d(x))$ in order to perform a change of
variables from the spatial coordinate $x$ to energy $\varepsilon$:
\begin{align}
  \label{eq:119}
   \kappa_{xy}^{\rm tr} &= -\frac {1} {\partial_x T} {\rm Tr}_+ \int_p    {\sf \Omega}^{\rm h}_{p_y p_x} \lim_{L\rightarrow \infty}\frac 1 L\int \text dx \,
                       \partial_x \left ( {\sf K}_d(x) \right )
                          \int \text d \varepsilon \,
                          \delta(\varepsilon-{\sf K}_d(x)) \,\varepsilon \,  \left ( n_{\rm B}(\varepsilon,T) \big |_{T(x)} + \tfrac 1 2 \right ) \Lambda(x) \nonumber \\
                  &= \frac {1} {\partial_x T} {\rm Tr}_+ \int_p    {\sf \Omega}^{\rm h}_{p_y p_x} \lim_{L\rightarrow \infty}\frac 1 L\int \text dx
                        \int \text d \varepsilon \, \partial_x
                    \Theta(\varepsilon-{\sf K}_d(x)) \,\varepsilon \,  \left ( n_{\rm B}(\varepsilon,T) \big |_{T(x)} + \tfrac 1 2 \right ) \Lambda(x) \nonumber \\
                  &= - \frac {1} {\partial_x T} {\rm Tr}_+ \int_p    {\sf \Omega}^{\rm h}_{p_y p_x} \lim_{L\rightarrow \infty}\frac 1 L\int \text dx
                        \int \text d \varepsilon \, \Theta(\varepsilon-{\sf K}_d(x)) \,\varepsilon \,  \partial_x  n_{\rm B}(\varepsilon,T) \big |_{T(x)}.
\end{align}
In going to the second line we used the chain rule, and in going to the third line we integrated by parts --
no boundary term is generated because of the cutoff function $\Lambda(x)$,
and the extra term proportional to $\partial_x \Lambda$ vanishes like $O(1/L)$.
\end{widetext}
We use finally that
\begin{align}
  \label{eq:120}
  \partial_x  n_{\rm B}(\varepsilon,T) \big |_{T(x)} &= - (\varepsilon/T(x)) \partial_x T(x) \partial_\varepsilon n _{\rm B}(\varepsilon,T) \big |_{T(x)},
\end{align}
and do not include $T(x)$ position dependence anywhere else,
and assume that $\partial_x T(x)$ 
is uniform within $I_x$ and zero outside, i.e.\ $\partial_x T(x)={\rm
  const}\neq0$ for $x\in I_x$ and $\partial_x T(x)=0$ for $x\notin I_x$. 
This is enough for the linear response regime up to $O(1/L)$ corrections thanks to the
decay of $g(x)$ over distances $O(L^0)$.
Since then $g(x)=1$ in the whole support of the integrand, one can
replace ${\sf K}_d(x)  \rightarrow {\sf K}_d$
as well as $\frac 1 L \int_{I_x} \text d x \rightarrow 1$, which yields exactly Eq.~\eqref{eq:117},
where we also used $  {\sf \Omega}^{\rm h}_{p_x p_y} = -   {\sf \Omega}^{\rm h}_{p_y p_x} $.

\section{The example of inhomogeneous elasticity with a time-reversal-breaking term}
\label{sec:example-inhom-elast}

\subsection{General}
\label{sec:general-1}

The Lagrangian density Eq.~\eqref{eq:98} is equivalent to the Hamiltonian density
\begin{equation}
  \label{eq:99}
\mathcal{H}_{\rm ph}(r)=\frac{1}{2\rho(r)}\Pi_i \Pi_i+\frac{c_{ij\mu\nu}(r)}{2}\partial_\mu u_i\partial_\nu u_j,
\end{equation}
upon introducing the conjugate lattice momentum
\begin{equation}
  \label{eq:100}
  \Pi_i=-\frac{i}{\hbar}\frac{\delta}{\delta u_i} - A_i[u],\quad
  A_i[u] \equiv\eta_{ij\mu\nu}(r)\partial^2_{\mu\nu}u_j .
\end{equation}
Thus the viscosity term introduces an effective vector potential which couples to the lattice momentum
and entails a nontrivial $[\Pi_i(r), \Pi_j(r')]$ commutator.
For convenience, we now define the ``reduced'' 
momentum ${\pi}_i(r)=\frac{\Pi_i(r)}{\rho(r)}$.
To connect with the general formulation of the main text, we define the 6-dimensional vector
\begin{equation}
  \label{eq:514}
  \Phi(r)=
  \begin{bmatrix} u_1(r), u_2(r), u_3(r), {\pi}_1(r), {\pi}_2(r), {\pi}_3(r) \end{bmatrix}^\top  ,
 \end{equation}
 and by comparing Eq.~\eqref{eq:99} and the commutation relations of the $u_i(r)$ and $\pi_j(r')$ fields
 to Eqs.~(\ref{eq:1},\ref{eq:2}) in the main text, one can readily identify $\ms \Gamma$ and $\ms H$.
Performing the Wigner transform to phase space coordinates,
 one then obtains the dynamical matrix $\ms K = \hbar\,\ms \Gamma \star \ms H$:
\begin{equation}
  \label{eq:101}
  {\sf \Gamma}=\frac{i}{\hbar\rho}
  \begin{bmatrix}
    0&1\\
    -1&  [\ms \Gamma_{22}]
  \end{bmatrix}, \; {\sf H}=\begin{bmatrix}
     [\ms H_{11}]&0\\
      0&\rho
      \end{bmatrix},
  \;   \ms K 
  = i \begin{bmatrix}
0 & 1\\
  [\ms K_{21}] &  [\ms K_{22}] 
  \end{bmatrix},
    \end{equation}
    \begin{widetext}
where each block is a $3 \times 3$ matrix indexed by $i,j=1..3$, and 
      \begin{align}
        [\ms H_{11}] &=  \frac{1}{4}\partial^2_{X_\mu X_\nu} c_{\mu\nu}+ \frac 1 {\hbar^2} c_{\mu\nu}p_\mu p_\nu , \nonumber\\
        [\ms \Gamma_{22}]  &= \frac{1}{2\hbar\rho}\left( \partial^2_{X_\mu X_\nu} \eta_{\mu\nu}
                             +2\eta_{\mu\nu} \left(\frac{\partial_{X_\mu}\rho \partial_{X_\nu}\rho }{\rho^2}
                             -\frac{\partial^2_{X_\mu X_\nu} \rho }{\rho} -\frac 2 {\hbar^2} p_\mu p_\nu\right)\right) ,  \nonumber\\
        [\ms K_{21}] &=-\frac{1}{\rho}\left(\frac14 \partial^2_{X_\mu X_\nu} c_{\mu\nu} +
   \frac 1 {\hbar^2}  c_{\mu\nu} p_\mu  p_\nu\right) - \upsilon^2\frac{
    c_{\mu\nu}}{4\rho}\left(\frac{\partial^2_{X_\mu X_\nu}\rho}{\rho} - 2
                  \frac{\partial_{X_\mu}\rho
                       \partial_{X_\nu}\rho}{\rho^2}\right) + \frac
                       { i \upsilon}{\hbar \rho} c_{\mu\nu}p_\nu \frac{\partial_{X_\mu}\rho}{\rho},\nonumber\\
        [\ms K_{22}]  &=\frac{1}{2\hbar\rho}\left(\partial^2_{X_\mu X_\nu} \eta_{\mu\nu}
                        +2\eta_{\mu\nu}\left(\frac{\partial_{X_\mu}\rho \partial_{X_\nu}\rho}{\rho^2}- \Big ( 1- \frac{\upsilon^2}2 \Big ) \frac{\partial^2_{X_\mu X_\nu} \rho}{\rho }
                        - \frac 2 {\hbar^2} p_\mu p_\nu\right)\right) + \frac {i \upsilon}{ \hbar \rho} \frac 2 {\hbar}  \eta_{\mu\nu}p_\nu \frac{\partial_{X_\mu}\rho}{\rho} .         \label{eq:124d}
      \end{align}
    \end{widetext}
Eqs.~(\ref{eq:101}--\ref{eq:124d}) involve parameters $\rho,
(c_{\mu\nu})_{ij}, (\eta_{\mu\nu})_{ij}$ whose dependence on position
$X$ can be chosen to be any arbitrary smooth functions of position
$X$, so that Eqs.~(\ref{eq:101}--\ref{eq:124d}) are the general expression, in phase space, of inhomogeneous (linear) elasticity with a Hall viscosity term.

     \subsection{Specific}
     \label{sec:specific}
     
     Now we focus on the special case of the simple spatial dependence
     from App.~\ref{sec:case-grav-field}.
     For this application we need only consider the spatially independent version of the above, Eq.~\eqref{eq:121} in the main text.
     We also choose the elasticity parameters to have the specific form Eq.~(\ref{eq:122b}), corresponding to
     the Lagrangian density 
     \begin{align}
       \label{eq:106}
       \mc L_{\rm ph} &= \frac{\rho}{2} \left(\partial_\tau \bs u \right)^2
      + \frac{1}{2} \Big ( - c_1 \bs u \cdot (\nabla^2 \bs u)
                         + c_2 (\bs \nabla_\perp \cdot \bs u_\perp)^2   \nonumber \\
                      & + c_3 (\partial_z u^z )^2  + c_4 (\bs \nabla_\perp u^z)^2 + c_5 (\partial_z \bs u_\perp )^2 \Big ) \nonumber \\
       &+ i \left ( \partial_\tau \bs u \times ( \eta_1 \nabla_\perp^2 + \eta_2 \partial^2_{zz} ) \bs u \right ) \cdot \mb{\hat z},
     \end{align}
     where $\bs \nabla_\perp  = (\partial_x,\partial_y,0)$ and $\bs u_\perp = (u^x, u^y,0)$.

  This problem can be solved analytically for arbitrary values of
  $\eta_1,\eta_2,c_a,a=1..5$. 
  In the following, we expand the solution in powers of the viscosity coefficients $\eta_{1,2}$,
    which we assume to be small with respect to $c_{1,2}\sqrt{\rho}$ -- this assumption is largely valid in known relevant cases \cite{ye2021}.
  We furthermore take $c_5=0$, which has little physical
     consequence but makes analytical expressions considerably
     shorter.
     We also define $\tilde c_a=c_a/\rho$ for $a=1..5$ and $\tilde \eta_b = \eta_b/\rho$ for $b=1,2$.
     
     We find the energy bands 
     \begin{align}
       \label{eq:125}
  {\sf K}^{\rm h}_d &= {\rm diag}\left ( \varepsilon_1 , -\varepsilon_1,
       \varepsilon_2, -\varepsilon_2, \varepsilon_3, -\varepsilon_3 \right ) ,\\
    \varepsilon_1 &= [\tilde c_1 \bs p^2]^{\frac 1 2}, \; \quad \varepsilon_2 = [ \tilde c_1 \bs p^2+ \tilde c_2 p_\perp^2]^{\frac 1 2}, \nonumber \\
         \varepsilon_3 &= [\tilde c_1 \bs p^2 + \tilde c_3 p_z^2+ \tilde c_4 p_\perp^2]^{\frac 1 2},  \nonumber 
     \end{align}
     where $p_\perp^2 = p_x^2+p_y^2$, and the ``normalized'' eigenvectors
     (normalized according to Eq.\eqref{eq:26})
     $\Psi_i, \Psi_i^*$ such that
     \begin{align}
       \label{eq:126}
       \ms S^{\rm h} &=
                       \begin{bmatrix}
                         \Psi_1 \;\vline\; \Psi_1^* \;\vline\;  \Psi_2 \;\vline\;  \Psi_2^* \;\vline\;  \Psi_3 \;\vline\;  \Psi_3^* 
                       \end{bmatrix}.
     \end{align}
     In writing the explicit expressions for the eigenvectors, we use the shorthand notations $\zeta_{\bs p}^{(\rm I)}\equiv \tilde\eta_1 p_\perp^2 + \tilde\eta_2\bs p^2$,
       $\zeta_{\bs p}^{(\rm II)}\equiv p_\perp^2(\tilde\eta_1 p_\perp^2 - \tilde\eta_2 p_z^2 )$
       and $\zeta_{\bs p}^{(\rm III)}\equiv \tilde\eta_1 p_\perp^4 - 2
       \tilde\eta_2 p_z^4 - \tilde\eta_2 p_z^2 p_\perp^2$, so that a choice of properly normalized eigenvectors can be
     \begin{widetext}
       \begin{align}
         \label{eq:128}
         \Psi_1 &= \frac 1 {\sqrt {2\rho}} \left [1+(p_y/p_x)^2 \right ]^{-\frac 1 2}
                  \left (  \frac{i}{\varepsilon_1}\frac{p_y}{p_x}  +  \frac{2 \zeta_{\bs p}^{(\rm I)}}{\tilde c_2 p_x^2}, - \frac{i}{\varepsilon_1} ,0,
        \frac{p_y}{p_x} - i \varepsilon_1 \frac{2 \zeta_{\bs p}^{(\rm I)}}{\tilde c_2 p_x^2} , - 1, 0 \right )^\top \quad + O(\eta^2),  \nonumber\\
         \Psi_2 &=\frac 1 {\sqrt {2\rho}} \left [1+(p_x/p_y)^2 \right ]^{-\frac 1 2}
                  \left ( \frac{i }{\varepsilon_2}\frac{p_x}{p_y}+ \frac{2 \zeta_{\bs p}^{(\rm I)}}{\tilde c_2 p_y^2}, \frac{i}{\varepsilon_2},0,
        \frac{p_x}{p_y} - i \varepsilon_2 \frac{2 \zeta_{\bs p}^{(\rm I)}}{\tilde c_2 p_y^2} ,1, 0 \right )^\top \quad + O(\eta^2),
       \end{align}
      \end{widetext}
      and $ \Psi_3 = \frac 1 {\sqrt {2\rho}} \left ( 0, 0, \frac{i}{\varepsilon_3} , 0, 0, 1 \right )^\top $.
      
  Here we provide $\Psi_1$ and $\Psi_2$ to linear order in
  $\eta_{1,2}$ only. 
     This is sufficient to compute the leading contribution to the magnetization ${\sf M}_{p_x p_y}$ and Berry curvature $\ms \Omega_{p_x p_y}$,
     which are odd under time reversal, and thus odd in powers of $\eta_{1,2}$.
     In particular $\Psi_3$ does not depend on $\eta_{1,2}$, therefore the magnetization and Berry curvature are zero in the third band.
     In the other two bands we find, explicitly,
     \begin{align}
       \label{eq:102}
       {\sf M}_{p_x p_y}^{(1)}&=\frac 1 { \varepsilon_1^2 p_\perp^2}  \tilde c_1 \zeta_{\bs p}^{(\rm III)}\quad +O(\eta^3) ,\nonumber\\
       {\sf M}_{p_x p_y}^{(2)}&= \frac 1 { \varepsilon_2^2 p_\perp^2 } \left ( \tilde c_2 \zeta_{\bs p}^{(\rm II)} + \tilde c_1 \zeta_{\bs p}^{(\rm III)} \right ) \quad +O(\eta^3) ,
     \end{align}
     \begin{widetext}
       \begin{align}
         \label{eq:123}
     {\ms \Omega}_{p_x p_y}^{(1)}&= \frac 1 { \tilde c_2 p_\perp^4 } \frac{\tilde c_1}{\varepsilon_1^3}
                                   \left ( \tilde c_2 p_\perp^2 \zeta_{\bs p}^{(\rm II)} + 2 \tilde c_2 \eta_1 p_\perp^4 p_z^2
                                   + 4 \tilde c_1 \bs p^2 \zeta_{\bs p}^{(\rm III)} \right ) \quad +O(\eta^3),\nonumber\\
          {\ms \Omega}_{p_x p_y}^{(2)}&= - \frac 1 { \tilde c_2 p_\perp^4 } \frac{1}{\varepsilon_2^3}
                                        \left ( 3 \tilde c_2^2  p_\perp^2 \zeta_{\bs p}^{(\rm II)} + 4 \tilde c_1^2 \bs p^2 \zeta_{\bs p}^{(\rm III)}
                                        + \tilde c_1 \tilde c_2 p_\perp^2 \left [ 2 p_z^2 (\tilde\eta_1 p_\perp^2 + \tilde\eta_2 p_z^2) + 7 \zeta_{\bs p}^{(\rm III)} \right ] \right ) \quad +O(\eta^3).
       \end{align}
     \end{widetext}
    
     We make a final remark about momentum integration.
     The integrand in Eqs.~(\ref{eq:111},\ref{eq:113}) does not vanish at large energies due to the zero-point contribution to the boson density
     (see details in Appendix~\ref{sec:discussion--currents} and discussion in Sec.~\ref{sec:energy-current}).
     While this does not have physical consequences for the total current $\mathcal J_{X_y}=\mathcal J_{X_y}^{(1)}+\mathcal J_{X_y}^{(2)}$,
     it makes both integrals diverge independently in a continuum model.
     Thus, when evaluating momentum integrals we restore a conventional Brillouin zone ${\rm BZ} \equiv [-\pi/\xi,\pi/\xi]^3$.
     
\end{document}